\documentclass[11pt]{article}
\usepackage{comment,soul}
\usepackage[dvipsnames]{xcolor} 
\usepackage[hidelinks]{hyperref}
\usepackage{enumitem}
\usepackage{amsmath,amssymb,epsf,cite,graphicx,subfigure}
\usepackage{amsmath,braket,tikzsymbols}
\usepackage[bbgreekl]{mathbbol}
\usepackage{comment}

\numberwithin{equation}{section}
\newcommand{\nn}{\nonumber}

\definecolor{airforceblue}{rgb}{0.36, 0.54, 0.66}
\newcommand{\beq}{\begin{equation}}
\newcommand{\eeq}{\end{equation}}

\usepackage[mathscr]{euscript}

\def\lie{\pounds}
\def\dow{\partial}
\def\Df{\mathrm{D}}
\def\df{\mathrm{d}}

\def\scX{\mathscr{X}}

\def\lb{\left(}
  \def\rb{\right)}
\def\half{\frac12}

\begin{document}
\baselineskip=15.5pt
\pagestyle{plain}
\setcounter{page}{1}

\begin{center}
{\LARGE \bf Fractons in curved space}
\vskip 1cm

\textbf{Akash Jain$^{1,2,a}$ and Kristan Jensen$^{3,b}$}

\vspace{0.5cm}

{\small ${}^1$Institute for Theoretical Physics, University of Amsterdam, 1090
  GL Amsterdam, The Netherlands \vspace{.3cm}}

{\small ${}^2$Dutch Institute for Emergent Phenomena, 1090 GL Amsterdam, The Netherlands \vspace{.3cm}}

{\small ${}^3$Department of Physics and Astronomy, University of Victoria, Victoria, BC V8W 3P6, Canada\\}

\vspace{0.3cm}

{\tt \small ${}^a$a.jain@uva.nl,}
{\tt  \small ${}^b$kristanj@uvic.ca \\}

\medskip

\end{center}

\vskip1cm

\begin{center}
{\bf Abstract}
\end{center}

We consistently couple simple continuum field theories with fracton excitations
to curved spacetime backgrounds. We consider homogeneous and
isotropic fracton field theories, with a conserved $U(1)$ charge and dipole
moment. Coupling to background fields allows us to consistently define a
stress-energy tensor for these theories and obtain the respective Ward
identities. Along the way, we find evidence for a mixed gauge-gravitational anomaly in the symmetric tensor gauge theory which naturally couples to conserved dipoles. Our results generalise to systems with arbitrarily higher conserved
moments, in particular, a conserved quadrupole moment.

\hspace{.3cm}

\newpage

\tableofcontents

\section{Introduction}

\label{sec:intro}

We are interested in quantum mechanical models with
fractons~\cite{Chamon:2004lew, 2011AnPhy.326..839B, Haah:2011drr, Vijay:2015mka,
  Nandkishore:2018sel}. These models describe exotic, at this time purely
hypothetical, phases of quantum matter displaying features that challenge our
usual notions of quantum field theory in the continuum limit. Perhaps the most
striking of these is the existence of finite-energy excitations with restricted
mobility: fractons are excitations which are ``pinned'' to a point, lineons are
able to move along a one-dimensional sublattice of a lattice model, etc. In some
instances, like the X-cube model~\cite{Vijay:2016phm},
they exhibit an ultraviolet-sensitive but non-extensive ground state
degeneracy. A priori, it is unclear how to describe these phenomena with
textbook quantum field theory and, indeed, much recent attention
(see~\cite{Seiberg:2020bhn,Seiberg:2020wsg} and follow-up work) has been devoted
to answering the question of how to carefully coarse-grain these models and what
are the rules of the game for their continuum limits.

Fortunately, the strange behaviour of these phases of matter is intimately tied
with their symmetries, which we are well-positioned to study in field
theory. Models of fracton order have exotic spacetime symmetries, like a
conserved dipole moment, or subsystem symmetry as in e.g.~\cite{Vijay:2016phm,
  Williamson:2016jiq, You:2018oai}. It is intuitively simple to understand how a
conserved dipole moment leads to ``fracton'' excitations. Namely, individual
charges can carry finite energy, but an isolated charge cannot move without
changing the dipole moment. The ultraviolet sensitive ground state degeneracies
are also tied to these symmetries. Take the X-cube model
of~\cite{Vijay:2016phm}, a theory of $\mathbb{Z}_2$ spins on a hypercubic
lattice. For each plane of the lattice, the Hamiltonian has a $\mathbb{Z}_2$
subsystem symmetry that flips all of the spins on that plane. The ground state
of that model is not invariant under this subsystem symmetry, so that there is a
large space of vacua generated by acting with the symmetry. The ground state
degeneracy is parametrically the volume of the symmetry group, which is
sensitive to the number of lattice sites in each direction.

The focus of this work is to better understand the spacetime symmetries of simple, continuum models of fractons. It is the first in a series of works whose broad goal is to study the role of these symmetries, as well as their spontaneous breaking, in interacting models of fractons and the ensuing implications for transport. Ultimately, we endeavour to find new, interacting, and soluble models of fractons, whose low-energy symmetry breaking pattern and careful quantisation thereby inform us as to what we might expect in the low-energy physics of these exotic theories. A simple but useful future application is to construct theories of transport, both at zero and at finite temperature, i.e. the hydrodynamic description of fracton models, which is strongly constrained by symmetry. (See~\cite{Gromov:2020yoc,Grosvenor:2021rrt,Glorioso:2021bif,Pena-Benitez:2021ipo} for earlier work on the hydrodynamics of fracton field theories.) We hope that the resulting theory of transport may prove useful in giving predictions which, perhaps, lead to the experimental discovery of models with fracton order.

Here we take the very first steps. As in standard Lorentz-invariant field
theory, we would like to couple the symmetry currents of these models to
external sources. That is, we would like to study fractons in curved space. This
task is non-trivial on account of exotic spacetime symmetries, which lead to a
somewhat intricate coupling to a background spacetime. We focus on what could be
described as the ``most symmetric'' fracton models, which are isotropic and
contain a conserved $U(1)$ charge and dipole moment. Our methods naturally
extend to field theories with conserved multipole moments, including those with
conserved dipole moment and quadrupole trace. This particular symmetry pattern
is perhaps the most experimentally relevant one on the market, given the
arguments that these symmetries approximately govern real-world systems
including vortices in superfluid helium~\cite{Nguyen:2020yve,Doshi:2020jso}, defects in $2+1$-dimensional elastic
media~\cite{2018PhRvL.120s5301P}, and the lowest Landau level of a quantum Hall
state~\cite{Du:2021pbc}. See also~\cite{sous2019,Sous:2020ypq} for other proposals to realize models with conserved dipole moment.

The output of our analysis is a systematic description of the symmetry currents
of these models, the algebra of local symmetries, and their Ward identities,
which are stable under radiative corrections in the absence of
anomalies. Indeed, while it seems unlikely to us that simple fracton models
(like the scalar theory of Pretko we review in Section~\ref{S:review}) possess
anomalies, we do note that the coupling to external fields allows for the future
classification and computation of perturbative anomalies, assuming they can
exist.

As a byproduct of our work, we find that the symmetric tensor gauge theory with
local dipole symmetry considered in~\cite{Pretko:2016kxt} cannot
be consistently coupled to curved space in a covariant way, on account of the fact that it does not possess a conserved and gauge-invariant stress tensor.\footnote{The tensor gauge theory was ``minimally coupled'' to a time-independent spatial metric in~\cite{Gromov:2017vir,Slagle:2018kqf}, and those authors found that the resulting model was gauge-non-invariant unless the spatial metric is one of constant curvature. Our result goes beyond those, insofar as we find that there is no curved space definition of the tensor gauge theory which is simultaneously gauge- and diffeomorphism-invariant.} This is reminiscent of a gauge-gravitational mixed anomaly,
in the sense that there is an obstruction to simultaneously maintaining gauge invariance and covariance. It remains to be seen if the
tensor gauge theory can be redefined in a way so as to maintain gauge-invariance
while at the expense of covariance, as for mixed gauge-gravitational anomalies in relativistic field theory. 

This work is a stepping stone. The next step is to pose, and solve, interacting
large $N$ fracton models. This will be done in~\cite{LargeN}, mostly using the
imaginary time formalism at finite temperature. At least in those models, one
can deduce the low-energy symmetry breaking pattern and accompanying Goldstone
effective theory. This effective description brings us most of the way to a
theory of transport. From there, it is straightforward to generalise the methods
of~\cite{Banerjee:2012iz,Jensen:2012jh,Crossley:2015evo,Haehl:2015uoc,Jensen:2017kzi}  (as well as e.g.~\cite{Jensen:2014ama,Geracie:2015xfa,Novak:2019wqg, deBoer:2020xlc, Armas:2020mpr} on constructing theories of transport in the non-relativistic setting) to obtain the dissipative
hydrodynamic description of these models, either from the point of view of
constitutive relations and conservation equations, or from the point of view of
a Keldysh effective field theory.

The remainder of this paper is organized as follows. In Section~\ref{S:review}
we briefly review field theories with conserved dipole moment. The heart of the
paper is Section~\ref{S:curved}, where we couple the symmetry currents of these
models to background fields. This coupling leads to a notion of general
covariance, and an infinite-dimensional algebra of charges generated by
diffeomorphisms and gauge transformations. We deduce this algebra in
Section~\ref{sec:first-order}, which, for technical reasons, is easiest to do
when working in an analogue of the first-order formulation of the background
spacetime. We discuss the extension of our results to even more exotic models
with conserved multipole moments in Section~\ref{sec:multipole-moments}, and
wrap up with a discussion in Section~\ref{S:discussion}. We relegate a few
technical computations to the Appendix.

\emph{Note}: While this paper was nearing completion, we became aware
of~\cite{Bidussi:2021}, whose authors also study the problem of putting models
with conserved dipole moment into curved spacetime. 

\section{Field theories with a conserved dipole moment}

\label{S:review}

Consider a rotationally and translationally invariant field theory with a
conserved $U(1)$ charge and dipole moment. We denote the Hamiltonian as
${\rm H}$, momenta as ${\rm P}_a$, angular momenta as ${\rm M}_{ab}$, $U(1)$
charge as ${\rm Q}$, and dipole moments as ${\rm D}_a$. The symmetry algebra of
charges is
\begin{align}
  \begin{split}
    [{\rm P}_a,{\rm D}_b]
    &= i\delta_{ab} {\rm Q}\,, \\
    [{\rm M}_{ab},{\rm D}_c]
    &= i(\delta_{ac}{\rm D}_b - \delta_{bc}{\rm D}_a)\,, \\
    [{\rm M}_{ab},{\rm P}_c]
    &= i(\delta_{ac}{\rm P}_b - \delta_{bc}{\rm P}_a)\,, \\
    [{\rm M}_{ab},{\rm M}_{cd}]
    &= i(\delta_{ac}{\rm M}_{bd}
    - \delta_{bc}{\rm M}_{ad} - \delta_{ad}{\rm M}_{bc} + \delta_{bd}{\rm M}_{ac})\,,
    \label{eq:dipole-algebra}
  \end{split}
\end{align}
with all other commutators vanishing.  The $U(1)$ charge $\rm Q$ appears as a
central extension.

The work of~\cite{Pretko:2018jbi} prescribed a systematic procedure to writing
down the action for a charged scalar field $\Phi$ that is invariant under these
symmetries. We review this construction in this Section. Take, for example, a
scalar field theory described by the action
\begin{align}
  S &= \int \df t \df^d x \Big( i\Phi^*\partial_t \Phi
      + \lambda\, \Df_{ij}(\Phi^*,\Phi^*)\Df^{ij}(\Phi,\Phi)
      - V(\Phi^*\Phi)\Big)\,,
      \label{eq:scalar-action}
\end{align}
where
$\Df_{ij}(\Phi,\Phi) = \Phi\partial_i\partial_j\Phi - \partial_i
\Phi\partial_j\Phi$. It is easy to see that this theory is invariant under
constant $U(1)$ rotations of the complex scalar field
$\Phi \to e^{i\Lambda}\Phi$. In fact, this theory has another invariance under
spatially linear $U(1)$ rotations $\Phi \to e^{i\psi_ix^i}\Phi$. This latter
symmetry leads to the conserved dipole moment. To wit, we can compute the
conserved charge density and flux associated with the global $U(1)$ symmetry of
the theory to be
\begin{equation}
  J^t = \Phi^*\Phi, \qquad
  J^i = \dow_j \Big(i\lambda\,\Df^{ij}(\Phi^*,\Phi^*) \Phi^2
  - i\lambda\,(\Phi^*)^2\Df^{ij}(\Phi,\Phi) \Big),
  \label{scalar-currents}
\end{equation}
satisfying $\dow_t J^t + \dow_i J^i = 0$ on the solutions of the equations of
motion. It is easy to see that the total charge defined as
\begin{equation}
  Q = \int \df^d x\,J^t,
\end{equation}
is conserved. However, the dipole moment defined as
\begin{equation}
  D^i = \int \df^d x\,x^i J^t,
\end{equation}
is also conserved. This is precisely the Noether current associated with the
linear $U(1)$ rotations. This conservation implies that the $U(1)$ flux $J^i$ in
eq. \eqref{scalar-currents} can be expressed as the divergence of a dipole flux
$J^{ij}$ as $J^i = \dow_jJ^{ij}$.
The conservation of $U(1)$ charge and dipole moment are then simultaneously
encoded in the Ward identity
\begin{equation}
\label{E:currentConservation}
\partial_{\mu} J^{\mu} = \partial_t J^t  + \partial_i \partial_j J^{ij} = 0\,.
\end{equation}
Note that only the symmetric part of $J^{ij}$ appears here. For the purposes of
better understanding the symmetries of the problem, we can then regard $J^{ij}$
as symmetric.\footnote{More precisely, since the antisymmetric part of $J^{ij}$
  drops out of the conservation equation, this antisymmetric part represents an
  ambiguity in the definition of the dipole current. In the language of high
  energy physics, we may consider an ``improved'' version of the dipole current
  whereby we redefine it to be symmetric. We can do this as long as the
  antisymmetric part of the original dipole current is a gauge-invariant
  operator.}

There are infinitely more terms that can be included into the action
\eqref{eq:scalar-action} consistent with the symmetries, like
$\partial_t \Phi^*\dow_t\Phi$,
$(\phi^*)^2 f(\Phi^*\Phi)\Df^i{}_{i}(\Phi,\Phi) + (\text{c.c.})$, etc. The crucial
point is that terms with spatial derivatives are strongly constrained by the
conserved dipole moment. In particular, the standard rotationally invariant term
$\dow_i\Phi^*\dow^i\Phi$ is forbidden.

The particular set of allowed terms can be understood by coupling to background fields. One introduces a field $A_t$ which couples to the charge density $J^t$, and a field $a_{ij}$ (with $a_{ij}=a_{ji}$) which couples to the dipole current $J^{ij}$, with
\begin{equation}
  \delta S = \int \df t \df^dx
  \left( J^t \delta A_t + \frac{1}{2}J^{ij} \delta a_{ij}\right)\,.
\end{equation}
The action is now a functional of the quantum field $\Phi$ and the background
fields $A_t$ and $a_{ij}$. The peculiar form of current
conservation~\eqref{E:currentConservation} then arises if we impose a symmetry
under
\begin{equation}
  \label{E:transfo1}
  \Phi \to e^{i \Lambda(t,\vec x)} \Phi\,, \qquad
  A_t \to A_t + \partial_t \Lambda(t,\vec x)\,, \qquad
  a_{ij} \to a_{ij} - 2\partial_i \partial_j \Lambda(t,\vec x)\,.
\end{equation}      
The absence of a vector gauge field $A_i$ implies that while there is a
covariant derivative with respect to time, i.e.
\begin{equation}
  \Df_t \Phi = \partial_t \Phi - i A_t \Phi\,,
\end{equation}
but no covariant derivative in spatial directions. Instead, the simplest
covariant object that acts with spatial derivatives on charged fields includes
two fields and two derivatives,
\begin{equation}
  \Df_{ij}(\Phi,\Phi) = \Phi \partial_i \partial_j \Phi
  - \partial_i \Phi \partial_j \Phi +\frac{i}{2} a_{ij}\Phi^2\,,
\end{equation}
which one can readily verify transforms covariantly under~\eqref{E:transfo1},
with $\Df_{ij}(\Phi,\Phi) \to e^{2i \Lambda} \Df_{ij}(\Phi,\Phi)$.  Acting on
two fields $\Phi_1$ and $\Phi_2$ with charges $q_1$ and $q_2$, there is a more
general expression
\begin{equation}
  \Df_{ij}(\Phi_1,\Phi_2)
  = \frac{1}{2}\left(
    \frac{q_1}{q_2}\Phi_1\partial_i\partial_j \Phi_2
    + \frac{q_2}{q_1} \Phi_2\partial_i \partial_j \Phi_1
    - \partial_i \Phi_1 \partial_j \Phi_2
    - \partial_j \Phi_1 \partial_i \Phi_2\right)
  + i \frac{q_1+q_2}{4}a_{ij}\Phi_1\Phi_2\,,
\end{equation}
which transforms as 
\begin{equation}
  \Df_{ij}(\Phi_1,\Phi_2) \to e^{i (q_1+q_2)\Lambda}\Df_{ij}(\Phi_1,\Phi_2)\,.
\end{equation}
For the quantum field theory of $\Phi$, it then follows that the simplest terms
in the effective action with spatial derivatives involve at least four powers of
$\Phi$. For this reason, one expects the model with spatial kinetic terms to be
strongly correlated, and indeed, in~\cite{LargeN} we find that this is the case in
soluble large $N$ generalizations of these theories.

There is also a simple free field theory of dynamical fields $A_t$ and $a_{ij}$, a symmetric tensor gauge theory analogous to pure
electromagnetism, first written down in~\cite{Pretko:2016kxt}. In this theory one identifies $(A_t,a_{ij})$ modulo the gauge symmetry. The gauge-invariant analogues of the electric and (the Hodge
dual of) magnetic fields in this theory are
\begin{align}
\label{E:dipoleEB}
  E_{ij} = -\partial_t a_{ij} - 2\partial_i \partial_j A_t\,, \qquad
  F_{ijk} = \partial_i a_{jk} - \partial_j a_{ik}\,,
\end{align}
and so it is easy to write down gauge-invariant actions, like
\begin{equation}
  \label{E:dipoleMaxwell}
  S = \int \df t \df^dx
  \left( \frac{\epsilon_0}{2}E_{ij}E^{ij}
    - \frac{1}{4\mu_0} F_{ijk}F^{ijk}\right)\,.
\end{equation}
As we will see, there is an obstruction to placing this theory in a general curved space.

\section{Coupling to curved space}

\label{S:curved}

In the previous Section, we studied how to write down simple continuum field
theories of a charged scalar field $\Phi$ with a conserved dipole moment. The
goal of the present section is to write such a theory covariantly, which will
allow us to place such a model into ``curved spacetime,'' coupling it to the
analogue of an external metric.

\subsection{Aristotelian background sources}

We are interested in physical systems that are invariant under spacetime
translations and spatial rotations, but with no boost symmetry -- Galilean or
Lorentz. Such systems naturally couple to the so-called Aristotelian
background\footnote{It is amusing to contrast the invocation of Aristotle with
  the more common naming convention in theoretical physics, where a result is named
  for the last author to discover it.}
sources~\cite{Novak:2019wqg,deBoer:2020xlc,Armas:2020mpr}.\footnote{These are a
  generalisation of the Newton-Cartan background sources that show up when
  coupling to Galilean (non-relativistic) field theories~\cite{Geracie:2014nka,Jensen:2014aia,Hartong:2014pma,Geracie:2015xfa}, but when no Galilean or Milne boost symmetry is
  imposed.} The sources consist of a clock-form $n_\mu$ and a degenerate
symmetric spatial metric tensor $h_{\mu\nu}$. Together, $n_\mu$ and $h_{\mu\nu}$
can be thought of as the analogue of the spacetime metric $g_{\mu\nu}$, but when
no Lorentz boost symmetry has been imposed to combine the space and time
components into a single object. Physically, $n_\mu$ couples to the energy
density $\epsilon^t$ and energy flux $\epsilon^i$ of the system, while
$h_{\mu\nu}$ couples to the momentum density $\pi^i$ and stress tensor
$\tau^{ij}$ respectively. Note that one of the components of $h_{\mu\nu}$ is not
independent due to the degeneracy condition. One typically also includes a gauge
field $A_\mu$ that couples to some conserved $U(1)$ particle-number/charge
density $J^t$ and the associated flux $J^i$ in the theory.

We denote the zero eigenvector of $h_{\mu\nu}$ by $v^\mu$, normalised as
$v^\mu n_\mu = 1$, such that $v^\mu h_{\mu\nu}=0$. This is to be understood as
the velocity of the preferred reference frame that is observing the physical
system under consideration. Using this, we can also define an inverse spatial
metric $h^{\mu\nu}$ satisfying $h^{\mu\nu}n_{\nu} = 0$ and
$h^{\mu\rho}h_{\nu\rho} = h^{\mu}_{\nu} = \delta^{\mu}_{\nu} - v^{\mu}n_{\nu}$.

An important aspect of Aristotelian spacetimes is that they come equipped with a
covariant derivative. Just like in general relativity, where the covariant
derivative is defined so that the metric tensor $g_{\mu\nu}$ is covariantly
constant, we can define an Aristotelian covariant derivative via the connection
\begin{equation}
  \Gamma^\lambda{}_{\mu\nu}
  = v^\lambda \partial_\mu n_\nu
  + \frac12 h^{\lambda\rho} \left(
    \partial_\mu h_{\nu\rho} + \partial_\nu h_{\mu\rho} - \partial_\rho h_{\mu\nu}
  \right).
  \label{eq:connection}
\end{equation}
The connection satisfies
\begin{equation}
\begin{gathered}
  \nabla_\lambda n_\nu = \nabla_\lambda h^{\mu\nu} = 0, \qquad
  \nabla_\lambda h_{\mu\nu} = - n_{(\mu} \lie_v h_{\nu)\lambda}, \qquad
  h_{\nu\mu}\nabla_\lambda v^\mu = \frac12 \lie_v h_{\nu\lambda}, 
  \\
  \Gamma^\mu_{\mu\nu} + F^n_{\nu\mu} v^\mu =
  \frac{1}{\sqrt{\gamma}}\partial_\nu\sqrt{\gamma}, \qquad
  T^\lambda{}_{\mu\nu}\equiv 2\Gamma^\lambda{}_{[\mu\nu]} = v^\lambda F^n_{\mu\nu},
\end{gathered}
\end{equation} 
where $\gamma = \det(n_\mu n_\nu + h_{\mu\nu})$,
$F^n_{\mu\nu} = \partial_\mu n_\nu - \partial_\nu n_\mu$, and $\pounds_X$ is the
Lie derivative along $X^{\mu}$. Round and square brackets indicate
symmetrisation and anti-symmetrisation over indices, with
$A_{(ab)}=\frac{1}{2}(A_{ab}+A_{ba})$ and
$A_{[ab]} =\frac{1}{2}(A_{ab}-A_{ba})$. A curious contrast compared to
Riemannian geometry is that the connection is torsional, with torsion
$T^{\lambda}{}_{\mu\nu}$; it is not possible to define a torsionless connection
that annihilates $n_\mu$. Note that this is purely a matter of definition and
has no physical relevance. One could also add a more general form of torsion to
this connection or use a different connection that annihilates $n_\mu$ and
$h^{\mu\nu}$, but we shall refrain from delving into these possibilities
here.\footnote{For completeness, we note that the most general form of the
  Aristotelian connection with the properties
  $\nabla_\lambda n_\mu = \nabla_\lambda h^{\mu\nu} = 0$ is given as
  \begin{equation}
    \Gamma^\lambda{}_{\mu\nu}
    = v^\lambda \partial_\mu n_\nu
    + \frac12 h^{\lambda\rho} \left(
      \partial_\mu h_{\nu\rho} + \partial_\nu h_{\mu\rho} - \partial_\rho h_{\mu\nu}
    \right)
    + n_{(\mu}\Omega_{\nu)\rho} h^{\lambda\rho}
    + \half \lb T^\lambda{}_{\mu\nu} - 2h_{\rho(\nu} T^\rho{}_{\mu)\sigma} h^{\sigma\lambda} \rb,
  \end{equation}
  where $T^\lambda_{\mu\nu} = h^\lambda_\rho\Gamma^\rho_{\mu\nu}$ (satisfying
  $n_\lambda T^\lambda_{\mu\nu}$) is the spatial torsion tensor and
  $\Omega_{\mu\nu} = 2h_{\lambda[\nu}\nabla_{\mu]} v^\lambda$ is the background
  frame vorticity tensor. \label{foot:torsion}} We can define the curvature
tensor as
\begin{equation}
  R^\lambda{}_{\rho\mu\nu}
  = \dow_\mu \Gamma^{\lambda}{}_{\nu\rho}
  - \dow_\nu \Gamma^{\lambda}{}_{\mu\rho}
  + \Gamma^\lambda{}_{\mu\sigma}\Gamma^\sigma{}_{\nu\rho}
  - \Gamma^\lambda{}_{\nu\sigma}\Gamma^\sigma{}_{\mu\rho}.
\end{equation}
Note that $n_\lambda R^\lambda{}_{\rho\mu\nu} = 0$.

By introducing an Aristotelian background, we can now take a non-relativistic
field theory and render it generally covariant, by coupling it to sources in
such a way as to be invariant under diffeomorphisms and gauge transformations.
Parametrising these symmetry transformations by $\scX=(\chi^\mu,\Lambda)$, their
infinitesimal action on the background fields is given as
\begin{equation}
  \begin{alignedat}{2}
    \delta_\scX n_\mu
    &= \lie_\chi n_\mu
    &&= \nabla_\mu(n_\lambda\chi^\lambda) + \chi^\lambda F^n_{\lambda\mu}\,, \\
    \delta_\scX h_{\mu\nu}
    &= \lie_\chi h_{\mu\nu}
    &&= \chi^\lambda \nabla_\lambda h_{\mu\nu}
    + 2h_{\lambda(\mu} \nabla_{\nu)}\chi^\lambda\,, \\
    \delta_\scX A_\mu
    &= \lie_\chi A_\mu + \dow_\mu\Lambda
    &&= \nabla_\mu(\Lambda + A_\mu\chi^\mu) + \chi^\lambda F_{\lambda\mu}\,.
  \end{alignedat}
  \label{eq:background-variations-diff}
\end{equation}
Here $F_{\mu\nu} = \dow_\mu A_\nu - \dow_\nu A_\mu$ is the $U(1)$ strength. The
action of the symmetry transformations on the derived fields $v^\mu$ and
$h^{\mu\nu}$ can also be obtained accordingly
$\delta_\scX v^\mu = \lie_\chi v^\mu$ and
$\delta_\scX h^{\mu\nu} = \lie_\chi h^{\mu\nu}$.

This coupling to background is rather trivial for an ``ordinary''
non-relativistic theory. All we have done is make precise what needs to be done
in order to write a translationally and rotationally invariant field theory in a
general set of coordinates. However, more structure is required for theories
that have a conserved dipole moment, where there is a richer interplay between
internal and spacetime symmetries.

\subsection{The dipole shift symmetry}

Before we adapt the Aristotelian backgrounds to account for physical systems
with conserved dipole moment, we need to pay a closer attention to the dipole
symmetry.  In Section~\ref{S:review}, we reviewed how, when coupling the density
$J^t$ and dipole flux $J^{ij}$ to the background fields $A_t$ and $a_{ij}$
respectively, the dipole moment conservation can be understood as the invariance
of the theory under a $U(1)$ transformation of the background fields:
$A_t \to A_t + \partial_t\Lambda$,
$a_{ij} \to a_{ij} - 2\partial_i\partial_j\Lambda$. Note that the symmetry acts
on $a_{ij}$ with a non-Leibniz differential operator.  In order to discuss the
algebra of charges generated by gauge transformations and diffeomorphisms, it is
convenient to instead realise this symmetry with linear differential
operators. This approach is also useful when coupling to a background.

To this end, we introduce a vector gauge field $A_i$ coupled to the flux $J^i$
and impose the usual invariance under
\begin{equation}
  A_t \to A_t + \partial_t\Lambda\,, \qquad
  A_i \to A_i + \partial_i\Lambda\,,
\end{equation}
leading to the regular $U(1)$ conservation law $\partial_{\mu}J^{\mu}= 0$. We
supplement it with an additional ``dipole shift symmetry'' given as
\begin{equation}
  A_i \to A_i + \psi_i, \qquad
  a_{ij} \to a_{ij} + \partial_i\psi_j + \partial_j\psi_i, 
\end{equation}
which imposes the constraint $J^i = \partial_j J^{ij}$. Together these relations
lead to the desired conservation equation
$\partial_t J^t + \partial_i\partial_j J^{ij} = 0$. Of course, we can entirely
``gauge fix'' the dipole shift symmetry by choosing $A_i = 0$, which forces us
to set $\psi_i = - \partial_i\Lambda$ and gives back our original $U(1)$
symmetry.

We already have a gauge field $A_\mu$ corresponding to a conserved $U(1)$ current
in the Aristotelian framework. To account for the conserved dipole moment, we
also need to introduce a degenerate symmetric spatial dipole gauge field
$a_{\mu\nu}$, constrained as $v^\mu a_{\mu\nu} = 0$. In the reference frame of
the background observer, when $v^\mu = \delta^\mu_t$, this reduces to the dipole
gauge field $a_{ij}$ discussed previously. The dipole shift transformations can
now be stated as
\begin{equation}
  A_\mu \to A_\mu + \psi_\mu, \qquad
  a_{\mu\nu} \to a_{\mu\nu}
  + h_{\mu}^\rho h_{\nu}^\sigma
  \lb \nabla_\rho \psi_\sigma + \nabla_\sigma\psi_\rho \rb,
\end{equation}
for some dipole shift parameter $\psi_\mu$ obeying $v^{\mu}\psi_{\mu}=0$. We
will study the consistency of this symmetry as a Lie algebra in the next
Subsection.

Note that the dipole source $a_{\mu\nu}$ cannot be used as a connection to
define ``dipole-covariant derivatives,'' because it is only sensitive to the
symmetric spatial derivative of $\psi_\mu$. However, we can define an object
that is nearly a dipole connection by combining $F_{\mu\nu}$ and $a_{\mu\nu}$ as
\begin{equation}
  A^\lambda{}_\mu = n_\mu v^\rho F_{\rho\sigma} h^{\sigma\lambda}
  + \half\lb h_{\mu}^\rho F_{\rho\sigma} h^{\sigma\lambda}
  + a_{\mu\sigma}h^{\sigma\lambda} \rb.
  \label{dipole-connection}
\end{equation}
Note that $n_\lambda A^\lambda{}_\mu = 0$.  It can be checked that this object
transforms as
\begin{equation}
  A^\lambda{}_\mu \to A^\lambda{}_\mu
  + \nabla_\mu \psi^\lambda
  + n_\mu \psi^\nu \nabla_\nu v^\lambda,
\end{equation}
where $\psi^\mu = h^{\mu\nu}\psi_\nu$. We can also define the ``dipole field strength''
\begin{equation}
\label{E:dipoleField}
  F^\lambda{}_{\mu\nu}
= \nabla_\mu A^\lambda{}_\nu - \nabla_\nu A^\lambda{}_\mu
  + F^n_{\mu\nu} v^\rho A^\lambda{}_\rho
  + 2n_{[\mu} A^\rho{}_{\nu]} \nabla_\rho v^\lambda\,,
\end{equation}
which transforms as
\begin{equation}
\label{E:dipoleFieldTransfo}
  F^\lambda{}_{\mu\nu}
  \to F^\lambda{}_{\mu\nu}
  + \lb R^{\lambda}{}_{\rho\mu\nu} + F^n_{\mu\nu} \nabla_\rho v^\lambda
  - 2 n_{[\mu}\nabla_{\nu]} \nabla_\rho v^\lambda \rb \psi^\rho.
\end{equation}

The ``dipole field strength'' $F^{\lambda}{}_{\mu\nu}$ is nothing more than the
curved space version of the dipole electric/magnetic fields discussed
in~\eqref{E:dipoleEB}. Note $F^{\lambda}{}_{\mu\nu}$ is not dipole-invariant in
a general background. As a result there is no way to define curved space
versions of the dipole electric and magnetic fields $E_{ij}$ and $B_{ijk}$ while
preserving the dipole symmetry. This presents an obstruction to coupling the
symmetric tensor gauge theory in~\eqref{E:dipoleMaxwell} to a curved spacetime,
which we discuss briefly at the end of this Section.

These definitions will be useful later when we compute the algebra of charges.
There, $A^\lambda{}_{\mu}$ and $F^\lambda{}_{\mu\nu}$ appear as the dipole
versions of $A_\mu$ and $F_{\mu\nu}$ respectively.

In is interesting to note that a priori $A^\lambda{}_\mu$ has $d(d+1)$
independent components on account of the condition
$n_\lambda A^\lambda{}_\mu = 0$, as we would expect for a connection for a
$d$-parameter symmetry transformation. However, in our case, it only has
$d(d+1)/2$ independent components in the form of
$a_{\mu\nu} = 2h_{\lambda(\mu}h^\rho_{\nu)} A^\lambda{}_{\rho}$. The remaining
$d(d+1)/2$ components are fixed in terms of the $U(1)$ field strength
$F_{\mu\nu}$ via the relation
$A^\lambda{}_{\mu} h_{\nu\lambda} - A^\lambda{}_{\nu} h_{\mu\lambda} =
F_{\mu\nu}$. This may seem like a technical aside, but there is some potentially
interesting physics here. In Section~\ref{S:review} we considered field theories
with a conserved dipole moment. In these models the elementary, bare,
excitations are $U(1)$ charges. Charges can acquire dipole moments through
quantum corrections, and perhaps there might be bound charge-anticharge states
with nonzero dipole moment. For such models the dipole density is not an
independent operator from the charge density. However one can envision effective
low-energy field theories with elementary \emph{dipoles} with an ``internal''
dipole density $\mathcal{D}^i$, so that the total dipole moment is of the form
\begin{equation}
	D^i = \int d^dx \left( J^t x^i + \mathcal{D}^i\right)\,.
\end{equation}
In such a model the dipole Ward identity would be modified to become
$J^i = \partial_t \mathcal{D}^i + \partial_j J^{ij}$, where the dipole current
$J^{ij}$ need no longer be symmetric. The ``internal'' dipole density
$\mathcal{D}^i$ and antisymmetric part of $J^{ij}$ have just the right number of
independent components to couple to the remaining $d(d+1)/2$ components of
$A^{\lambda}{}_{\mu}$. Indeed, it is easy to see that this is the correct
interpretation for the remaining components of $A^{\lambda}{}_{\mu}$, i.e. if
they were present, they would couple to precisely such an ``internal'' dipole
density and antisymmetric dipole current.

\subsection{Conserved currents and Ward identities}
\label{S:ward}


We define the symmetry currents through the variation of the generating
functional $W = -i \ln Z$ with respect to background fields:\footnote{We have
  decided to couple $J^{\mu\nu}$ to the full connection $A^\lambda{}_{\mu}$
  instead of the spatial connection $a_{\mu\nu}$. This is purely a matter of
  definition because
  $J^{\mu\nu}h_{\nu\lambda} \delta A^\lambda{}_\mu = J^{\mu\nu} v^\rho
  F_{\rho\nu} \delta n_\mu - J^{\mu\rho} A^\nu_{~\rho} \delta h_{\mu\nu} + \half
  J^{\mu\nu} \delta a_{\mu\nu}$. This should be thought of as a convenient
  definition of other conserved currents in the presence of a dipole symmetry,
  so that they have ``nice'' transformations properties under the dipole shift
  symmetry.}
\begin{equation}
  \delta W
  = \int \df ^{d+1} x \sqrt{\gamma}\lb
  - \epsilon^\mu \delta n_\mu
  + \lb v^{(\mu} \pi^{\nu)} + \half \tau^{\mu\nu} \rb \delta h_{\mu\nu}
  + J^\mu \delta A_\mu
  + J^{\mu}{}_\lambda\delta A^\lambda{}_{\mu}
  \rb,
  \label{eq:actionVariation}
\end{equation}
where $\gamma = \det(n_\mu n_\nu + h_{\mu\nu})$. Here $\epsilon^{\mu}$ is the
energy current, $\pi^{\mu}$ the momentum current satisfying
$\pi^{\mu}n_{\mu} = 0$, $\tau^{\mu\nu}$ the spatial stress tensor satisfying
$\tau^{\mu\nu}n_{\nu} = 0$, $J^{\mu}$ the $U(1)$ current, and
$J^{\mu}{}_{\lambda}$ the dipole current. This expression should be understood
as the ``covariant definition'' of the conserved currents in curved
spacetime. Explicitly picking coordinates $x^{\mu} = (t,x^i)$, where we only
mandate $v^t \neq 0$, the various components of the currents read
\begin{gather}
  \epsilon^\mu =
  \begin{pmatrix}
    \epsilon \\ \epsilon^i
  \end{pmatrix}, \qquad
  \pi_\mu =
  \begin{pmatrix}
    - v^k \pi_k/v^t \\ \pi_i
  \end{pmatrix}, \qquad
  \tau^{\mu\nu} =
  \begin{pmatrix}
   n_k n_l\tau^{kl}/n_t^2 & - n_k \tau^{kj}/n_t \\
   - n_k \tau^{ik}/n_t & \tau^{ij}
 \end{pmatrix}, \nn\\
 J^\mu =
 \begin{pmatrix}
   J^t\\ J^i
 \end{pmatrix}~, \qquad
 J^{\mu\nu} =
 \begin{pmatrix}
   n_k n_lJ^{kl}/n_t^2 & - n_kJ^{kj}/n_t \\
   - n_kJ^{ik}/n_t & J^{ij}
 \end{pmatrix},
 \label{eq:cov-operators}
\end{gather}
where indices have been raised and lowered using the spatial metric $h_{\mu\nu}$
and its ``inverse'' $h^{\mu\nu}$.  These expressions satisfy the various
identities $\pi_\mu v^\mu = \tau^{\mu\nu} n_\nu = J^{\mu\nu} n_\nu = 0$,
$\tau^{[\mu\nu]} = 0$, and $J^{[\mu\nu]} = 0$.

We have in mind a dipole-symmetric field theory coupled to a background
spacetime in such a way as to be invariant under diffeomorphisms, $U(1)$ gauge
transformations, and dipole transformations, acting on both the quantum and
background fields. More precisely, in the absence of anomalies, the generating
functional $W$ of the theory is required to be invariant under these
transformation, leading to the Ward identities. Let $\chi^{\mu}$ denote an
infinitesimal diffeomorphism, $\Lambda$ an infinitesimal gauge transformation,
and $\psi_{\mu}$ an infinitesimal dipole transformation. Collectively denoting
them as $\hat\scX = (\chi^\mu,\Lambda,\psi_\mu)$, the action of these
transformations on the background fields is given by
\begin{align}
\begin{split}
  \delta_{\hat\scX} n_\mu
  &= \lie_\chi n_\mu\,, 
  \\
  \delta_{\hat\scX} h_{\mu\nu}
  &= \lie_\chi h_{\mu\nu}\,,
  \\
  \delta_{\hat\scX} A_\mu
  &= \lie_\chi A_\mu + \dow_\mu\Lambda + \psi_\mu\,,
  \\
  \delta_{\hat\scX} A^\lambda{}_\mu
  &= \lie_\chi A^\lambda{}_\mu
    + \nabla_\mu \psi^\lambda
    + n_\mu \psi^\nu \nabla_\nu v^\lambda\,.
     \label{eq:background-variations}
\end{split}
\end{align}
By assumption the generating functional is invariant under these
transformations, i.e. $\delta_{\hat\scX}W = 0$. This variation is given by
plugging the symmetry variation in eq.~\eqref{eq:background-variations} into
$\delta W$ in eq.~\eqref{eq:actionVariation}. Imposing that this symmetry
variation vanishes and integrating by parts, we find the following Ward
identities
\begin{align}
\begin{split}
  \nabla'_\mu \epsilon^\mu
  &= - v^\mu f_\mu
  - \lb \tau^{\mu\nu} + \tau^{\mu\nu}_{\text{d}} \rb
  h_{\lambda\nu}\nabla_{\mu} v^\lambda\,,
  \\
  \nabla'_\mu\lb v^\mu \pi^\nu + \tau^{\mu\nu}
  + \tau^{\mu\nu}_{\text{d}}\rb
  &= h^{\nu\mu} f_\mu
  - \pi_\mu h^{\nu\lambda} \nabla_\lambda v^\mu\,,
  \\
  \nabla'_\mu J^\mu
  &= 0\,,
  \\
  \nabla'_\mu J^{\mu\nu}
  &= h^\nu_\mu J^\mu\,,
  \label{eq:WardIdentity}
\end{split}
\end{align}
where $\nabla'_\mu = \nabla_\mu + F^n_{\mu\nu}v^\nu$ and
\begin{align}
  f_\mu
  &= -F^n_{\mu\nu} \epsilon^\nu - h_{\mu\lambda}A^{\lambda}{}_\nu J^\nu 
  + F^\lambda{}_{\mu\nu} h_{\rho\lambda} J^{\nu\rho}
  - n_\mu A^{\lambda}{}_\rho\, J^{\rho}{}_\nu \nabla_{\lambda} v^\nu, \nn\\
  \tau^{\mu\nu}_{\text{d}}
  &= - A^\mu{}_\rho J^{\rho\nu}.
    \label{eq:force-density}
\end{align}
The $f_{\mu}$ contributions to the right-hand-side of the energy and momentum
conservation Ward identities correspond to the power-force density due to the
background fields sourcing energy and momentum, analogous to the familiar Joule
heating term in $F_{\mu\nu}J^{\nu}$ in the stress tensor Ward identity of a
relativistic field theory. The terms involving $\nabla_\mu v^\nu$, on the other
hand, can be thought of as pseudo-power and pseudo-force contributions due to
the background observed not being inertial. On the other hand,
$\tau^{\mu\nu}_{\text{d}}$ is an effective contribution to the stress tensor due
to dipole sources.

The Ward identities are manifestly covariant under diffeomorphisms, with the
background fields and currents transforming as tensors, and are invariant under
gauge transformations. However, the behaviour of various fields under dipole
transformations is not so obvious. To derive the transformation laws of the
currents, we use the dipole symmetry to equate the generating functional $W$
with its value evaluated on a dipole-transformed background
\begin{equation}
  W[n_{\mu},h_{\nu\rho};A_{\mu},A^{\lambda}{}_{\mu}]
  = W[n_{\mu},h_{\nu\rho};~A_{\mu}+\psi_{\mu}~,~
  A^{\lambda}{}_{\mu} + \nabla_{\mu}\psi^{\lambda}
  + n_\mu \psi^\nu \nabla_\nu
  v^\lambda~]\,. 
\end{equation}
and then take a variation of this identity with respect to the various
background fields. The variations of the left-hand-side give the currents, and
those of the right-hand-side can be expressed in terms of the dipole-transformed
currents, which follow from the variations of with respect to the transformed
background fields. For infinitesimal $\psi_{\mu}$, this amounts to setting the
second variation $\delta (\delta_{\psi}W) $ to vanish. This second variation
reads
\begin{align}
\nn
  \delta(\delta_\psi W)
  &= \int \df^{d+1}x \sqrt{\gamma}\Big(
  - \delta_\psi\epsilon^\mu \delta n_\mu
  + \lb v^\mu \delta_\psi\pi^\nu + \half \delta_\psi\tau^{\mu\nu} \rb \delta
    h_{\mu\nu} 
    + \delta_\psi J^\mu \delta A_\mu
    + \delta_\psi J^{\mu}{}_\lambda \delta A^\lambda_{\mu}
    \\
  &\qquad \qquad\qquad\qquad
    + J^\mu \psi^\sigma \delta h_{\sigma\mu}
    + \lb 2J^{\mu(\rho}\psi^{\sigma)} - J^{\rho\sigma} \psi^\mu \rb
    \half\lie_v h_{\rho\sigma} \delta n_\mu 
    \\
    \nn
  &\qquad\qquad \qquad\qquad\qquad
    - J^{\mu\sigma} v^\rho F^n_{\mu\nu} \psi^\nu \delta h_{\sigma\rho} 
    + \half J^{\mu\nu} \psi^\lambda \nabla_\lambda\delta h_{\mu\nu}
  \Big)\,,
\end{align}
where we have used the dipole
Ward identity. In deriving this we have used the variation of the connection
\begin{align}
\begin{split}
  \delta \Gamma^\lambda{}_{\mu\nu}
  &= v^\lambda \nabla_\mu \delta n_\nu
    + \frac12 h^{\lambda\rho} \left(
    \nabla_\mu \delta h_{\nu\rho} + \nabla_\nu \delta h_{\mu\rho}
    - \nabla_\rho \delta h_{\mu\nu}
  \right) \\
  &\qquad
    + \half h^{\lambda\rho} (\lie_v h_{\mu\nu} )\delta n_\rho
    - \frac12 h^{\lambda\sigma} \lb v^\rho F^n_{\mu\nu} \delta h_{\sigma\rho}
    + 2v^\rho F^n_{\sigma(\mu} \delta h_{\nu)\rho} \rb\,.
  \end{split}
    \label{eq:connection-variation}
\end{align}
Reading off the contributions coming from individual background field
variations, we infer that the currents shift under dipole transformations as
\begin{align}
\begin{split}
  \epsilon^\mu
  &\to \epsilon^\mu
    + \lb 2J^{\mu(\rho}\psi^{\sigma)} - J^{\rho\sigma} \psi^\mu \rb
    \half\lie_v h_{\rho\sigma}\,,
    \\
  \pi^\mu
  &\to \pi^\mu - (J^\nu n_\nu) \psi^\mu
    + J^{\rho\mu} F^n_{\rho\sigma} \psi^\sigma\,,
    \\
  \tau^{\mu\nu}
  &\to \tau^{\mu\nu} - 2J^\lambda h_\lambda^{(\mu} \psi^{\nu)}
  + \nabla'_\lambda(\psi^\lambda J^{\mu\nu} )\,.
  \label{eq:dipole-var-currents}
\end{split}
\end{align}
while $J^\mu$ and $J^{\mu\nu}$ are invariant. One can explicitly see that this
is a symmetry of the Ward identities.

It is interesting to note that the energy density and flux contained in
$\epsilon^\mu$ are invariant under dipole shifts in flat space, but the same is
not true for the momentum density $\pi^\mu$ or the stress tensor
$\tau^{\mu\nu}$. One physical implication of this fact is that a relativistic
field theory cannot have a conserved dipole moment, because momentum density is
equated to energy flux by Lorentz boost symmetry. The same is true for a
Galilean theory, where the Galilean boost symmetry equates momentum density to
charge (particle number) flux. These transformation properties of conserved
currents can also be used to diagnose the consistency of seemingly
dipole-symmetric field theories written out in flat space. We shall illustrate
this in the next subsection for the symmetric tensor gauge theory given in
Eq.~\eqref{E:dipoleMaxwell}.

\subsection{Dipole-symmetric field theory in curved space}

We can use the Aristotelian background sources to write down the covariant 
version of the field theories mentioned in Section~\ref{S:review}. Firstly, we
note that we can define a gauge covariant derivative for the scalar field $\Phi$
with charge $q$ as
\begin{equation}
  \Df_\mu\Phi = \dow_\mu\Phi - iqA_\mu\Phi,
\end{equation}
however, it is not invariant under the dipole shift symmetry:
$\Df_\mu\Phi \to \Df_\mu\Phi - iq\psi_\mu\Phi$. The ``time-component'' of the
covariant derivative $v^\mu\Df_\mu\Phi$ is still dipole-invariant, but for the
spatial part we instead need to define a double derivative operator
\begin{equation}
  \Df_{\mu\nu}(\Phi,\Phi)
  = h_{(\mu}^\rho h_{\nu)}^\sigma \left(
    \Phi\,\Df_\rho \Df_\sigma\Phi
    - \Df_\rho\Phi\,\Df_\sigma\Phi
  \right)
  + \frac{iq}{2} a_{\mu\nu}\Phi^2.
\end{equation}
With this, the original scalar field theory may be written covariantly as
\begin{equation}
  S = \int \df^{d+1}x\sqrt{\gamma} \Big(
  i\Phi^* v^\mu \Df_\mu\Phi
  + \lambda h^{\mu\rho} h^{\nu\sigma} \Df_{\mu\nu}(\Phi^*,\Phi^*)
  \Df_{\rho\sigma}(\Phi,\Phi)
  - V(\Phi^*\Phi)
  \Big)\,.
  \label{eq:covariant-Phi}
\end{equation}
Now that we have a theory coupled to background sources, we can read off all the
conserved currents by varying with respect to these. We find
\begin{align}
\begin{split}
  \epsilon^\mu
  &= -v^\mu \Big(
    \lambda h^{\lambda\rho} h^{\tau\sigma} \Df_{\lambda\tau}(\Phi^*,\Phi^*)
    \Df_{\rho\sigma}(\Phi,\Phi)
    - V(\Phi^*\Phi)
    \Big)\,,
    \\
  &\qquad \qquad
    + 2\lambda \Big(
    \Df^{\mu\rho}(\Phi^*,\Phi^*)
    \lb \Phi\,v^\nu\Df_{(\nu} \Df_{\rho)}\Phi
    - v^\nu\Df_\nu\Phi\,\Df_\rho\Phi \rb
    + \text{c.c.}
    \Big)
    \\
  &\qquad \qquad \qquad 
    + \Big( iq \lambda \Df^{\mu\nu}(\Phi^*,\Phi^*) \Phi^2
    - iq \lambda \Phi^2 \Df^{\mu\nu}(\Phi,\Phi)
    \Big) v^\rho F_{\rho\nu} \,,
    \\
  \pi^\mu
  &= - i h^{\mu\nu} \Phi^*\Df_\nu\Phi\,,
  \\
  \tau^{\mu\nu} + \tau^{\mu\nu}_{\text{d}}
  &= h^{\mu\nu} \Big( i\Phi^* v^\nu \Df_\nu\Phi
    + \lambda h^{\lambda\rho} h^{\tau\sigma} \Df_{\lambda\tau}(\Phi^*,\Phi^*)
    \Df_{\rho\sigma}(\Phi,\Phi)
    - V(\Phi^*\Phi)
    \Big) \,,
    \\
  &\qquad \qquad
    - 4\lambda h_{\rho\sigma} 
    \Df^{\rho(\mu}(\Phi^*,\Phi^*)\Df^{\nu)\sigma}(\Phi,\Phi) 
    \\
  &\qquad \qquad \qquad 
    + \Big( iq \lambda \Df^{\mu\rho}(\Phi^*,\Phi^*) \Phi^2
    - iq \lambda \Phi^2 \Df^{\mu\rho}(\Phi,\Phi) \Big) A^{\nu}_{~\rho}\,,
    \\
  J^\mu
  &= q\Phi^*\Phi v^\mu
    + \Df_\nu \Big( iq \lambda \Df^{\mu\nu}(\Phi^*,\Phi^*) \Phi^2
  - iq \lambda \Phi^2 \Df^{\mu\nu}(\Phi,\Phi) 
          \Big)\,,
          \\
  J^{\mu\nu}
  &= iq \lambda \Df^{\mu\nu}(\Phi^*,\Phi^*) \Phi^2
    - iq \lambda \Phi^2 \Df^{\mu\nu}(\Phi,\Phi)\,.
\end{split}
\end{align}

What of the symmetric tensor gauge theory~\eqref{E:dipoleMaxwell}? The curved
space versions of the electric field $E_{ij} = 2F^i{}_{jt}$ and (the Hodge dual
of) magnetic fields $F_{ijk} = 2F^{k}{}_{ij}$ are encoded in the same object
$F^{\lambda}{}_{\mu\nu}$ defined in Eq.~\eqref{E:dipoleField}. However, as we
saw in Eq.~\eqref{E:dipoleFieldTransfo}, this object is not dipole-invariant in
a general spacetime. While we cannot conclusively rule out the possibility that
there is another dipole- and gauge-invariant curved space tensor that reduces to
$E_{ij}$ and $F_{ijk}$ in flat space, we have worked rather hard to find such an
object without success. As such, we are inclined to take this non-invariance
seriously and arrive at the conclusion that there is an obstruction to defining
a dipole-symmetric version of the electric and magnetic fields in a general
curved space.\footnote{There can, of course, be special curved backgrounds to
  which the symmetric tensor gauge theory can be coupled. A notable example
  comes from the work of~\cite{Gromov:2017vir,Slagle:2018kqf}. Consider a background spacetime
  where time is absolute and the spatial metric is time-independent, i.e.
  $\nabla_\mu v^\nu = \half h^{\nu\rho}h_{\mu}^\sigma\lie_v h_{\rho\sigma} = 0$,
  and whose spatial part is an Einstein manifold, i.e.
  $R^\lambda{}_{\rho\mu\nu} = \frac{1}{L^2} (h^\lambda_\mu h_{\rho\nu} -
  h^\lambda_\nu h_{\rho\mu})$. For such background spacetimes, we find
  $F^\lambda{}_{\mu\nu} \to F^\lambda{}_{\mu\nu} + \frac{1}{L^2} (h^\lambda_\mu
  \psi_{\nu} - h^\lambda_\nu \psi_{\mu})$. Consequently, the curved space
  electric field $F^\lambda{}_{\nu\rho}v^\rho$ and the ``symmetric'' magnetic
  field
  $F^\lambda{}_{\mu\nu} + \frac{2}{d-1} F^{\rho}{}_{\rho[\mu} h_{\nu]}^\lambda$
  are dipole-symmetric, and can be used to define a covariant version of the
  symmetric tensor gauge theory in Eq.~\eqref{E:dipoleMaxwell}. This conclusion
  is consistent with that of~\cite{Slagle:2018kqf}.} Because the dipole symmetry
is really a part of the gauge symmetry of this model, rather than a global
symmetry, this renders the symmetric tensor gauge theory inconsistent in a
general background.


This obstruction has a simple explanation in the flat space theory.  Consider
gauge fixing the dipole shift symmetry by setting $A_i = 0$, which fuses the
dipole transformation to $U(1)$ gauge transformations as
$\psi_i = -\dow_i\Lambda$. Then, using the transformation rules in
Eq. \eqref{eq:dipole-var-currents}, we infer that the energy current is gauge-invariant, but the momentum density $\pi^i$ and spatial stress tensor $\tau^{ij}$ 
are no longer gauge invariant. In fact, in flat space, they ought to transform as
$\pi^i \to \pi^i + J^t\dow^i\Lambda$ and $\tau^{ij} \to \tau^{ij} + 2 \partial_k J^{k(i)}\partial^j \Lambda - \partial_k (\partial^k \Lambda J^{ij})$. However, the equations of motion set
$J^t$  and $J^{ij}$ to zero. Consequently, the full stress tensor ought to be dipole-invariant
on-shell. Relatedly, the spatial stress tensor $\tau^{ij}$ ought to be symmetric on-shell. Interestingly, we find that this is not the case for the symmetric tensor gauge theory given by the action in
Eq.~\eqref{E:dipoleMaxwell}: the momentum current and spatial stress tensor are not gauge-invariant, nor is the spatial stress tensor symmetric.

Note that the canonical momentum density, the Noether density that generates
spatial translations, is already gauge-non-invariant in ordinary Maxwell
theory. But one can still define an improved gauge-invariant momentum density as
the generator of a spatial translation together with an appropriate gauge
transformation. Consider a symmetry transformation involving a spatial
translation along a constant translation parameter $\chi^i$ and a gauge
transformation $\Lambda = -\chi^i\lambda_i$, for some quantity $\lambda_i$ that
depends on the gauge fields. The improved momentum density of Maxwell theory is
\begin{equation}
  \pi_i = \frac{1}{\mu_0} \lb \dow_i A_j - \dow_j\lambda_i \rb E^j\,,
\end{equation}
which can be rendered gauge-invariant by setting $\lambda_i = A_i$, leading to
the standard expression for momentum density as $\pi_i = \frac{1}{\mu_0}
F_{ij}E^j$. Applying the same procedure to the symmetric tensor gauge theory in
Eq.~\eqref{E:dipoleMaxwell}, we get the momentum density
\begin{equation}
  \pi_i = \epsilon_0\lb \dow_i a_{jk} + 2\dow_j\dow_k\lambda_i \rb E^{jk}.
\end{equation}
This expression cannot be made gauge-invariant for any choice of
$\lambda_i$, so we may as well set $\lambda_i = 0$. 
We have similarly computed the energy current and spatial stress tensor for the theory~\eqref{E:dipoleMaxwell}. There is a gauge-invariant improved energy current, but the spatial stress tensor is not gauge-invariant either. We find
\begin{align}
  \tau^{ij}
  &= - \epsilon_0 E^{ik} E^j{}_{k}
  + \frac{1}{\mu_0} F^{ikl} F^j{}_{kl}
  + \frac12\delta^{ij}\lb
    \epsilon_0 E_{kl}E^{kl} - \frac{1}{2\mu_0} F_{klm}F^{klm}\rb \nn\\
  &\qquad
    - \epsilon_0 E^{ik} \dow_t \lb a^j{}_{k} + 2 \dow_k\lambda^j \rb
    - 2\epsilon_0 \dow_k E^{ki} \lb \dow^j A_t - \dow_t\lambda^j \rb
    + \frac{1}{\mu_0} F^{ikl} \dow_k \lb a^j{}_{l} 
    +  2\dow_l\lambda^j \rb,
\end{align}
which is neither gauge-invariant nor symmetric for any choice of $\lambda_i$.

Suppose, now, that we couple the flat space theory to linearised
perturbations of the background spacetime. On account of the non-symmetric spatial stress tensor, in order to maintain diffeomorphism invariance, this can be done only by coupling to a background in the first-order formalism, where we decompose the spatial metric into a spatial vielbein as $h_{\mu\nu} = \delta_{ab} e^a_{\mu}e^b_{\nu}$, with $a=1,2,\ldots,d$. This is done with a perturbation in
the action,
\begin{equation}
  \delta S = \int \df t \df^dx \left(-\epsilon^t \delta n_t - \epsilon^i \delta n_i +  \pi^i \delta e^i_t + \tau^{ij}\delta e^i_{j} \right)\,.
\end{equation}
This linearised coupling is diffeomorphism invariant on account of the conservation equations
\beq
	\partial_{\mu}\epsilon^{\mu} = 0\,, \qquad \partial_t \pi^i + \partial_j \tau^{ij} = 0\,,
\eeq
however it is not gauge-invariant, rendering the model
inconsistent. We therefore conclude that the symmetric tensor gauge theory
cannot be consistently coupled to curved spacetime while preserving
diffeomorphism invariance. This is reminiscent of a mixed gauge-gravitational
anomaly. We however hesitate to use that term just yet. In relativistic field
theories with mixed flavor-gravitational anomalies, one can redefine the theory
in such a way as to be either non-invariant under flavor transformations or
under diffeomorphisms. Here, it is not yet clear if one can redefine the tensor
gauge theory in such a way as to be gauge-invariant, but non-covariant in curved
space.

However, if charged matter coupled to the gauge fields condenses so that the
tensor gauge theory is in a Higgs phase, then the massive gauge theory may be
placed in curved spacetime. Consider the following invariant action where we
also introduce a charged scalar $\Phi$ with action
\begin{equation}
  S = \int\df^{d+1}x\,
  h_{\lambda\tau} h^{\nu\sigma}
  {\cal F}^\lambda{}_{\mu\nu} {\cal F}^{\tau}{}_{\rho\sigma}
  \lb 2\epsilon_0 v^\mu v^\rho
  - \frac{1}{\mu_0} h^{\mu\rho} \rb.
\end{equation}
We have defined the dipole-invariant combination of the dipole field strength
$F^\lambda{}_{\mu\nu}$ and the charged scalar field $\Phi$ via
\begin{equation}
  {\cal F}^\lambda{}_{\mu\nu} = \Phi^*\Phi F^\lambda{}_{\mu\nu}
  - \frac{i}{q}\Phi^*\Df_\rho\Phi\, h^{\rho\sigma}
  \lb R^{\lambda}{}_{\sigma\mu\nu} + F^n_{\mu\nu} \nabla_\sigma v^\lambda
  - 2 n_{[\mu}\nabla_{\nu]} \nabla_\sigma v^\lambda \rb.
\end{equation}
Of course, we can also add the usual dipole-invariant $\Phi$ terms to the action
from the theory in Eq.~\eqref{eq:covariant-Phi}. Unlike the theory in
Eq.~\eqref{E:dipoleMaxwell}, however, in the flat space limit the $E_{ij}E^{ij}$
and $F_{ijk}F^{ijk}$ terms in this theory come coupled to a factor of
$\Phi^*\Phi$.  Assuming that the dynamics allow $\Phi$ to condense, this model
has a Higgs phase whose low-energy description ought to be a massive version of
the symmetric gauge theory~\eqref{E:dipoleMaxwell} coupled to the phase of
$\Phi$. 

\section{First order formulation and dipole symmetry algebra}

\label{sec:first-order}

In the previous Section, we have worked with the second-order formulation of
Aristotelian geometries, where the curved spacetime background is captured by a
clock-form $n_\mu$ and a spatial metric $h_{\mu\nu}$. While this formulation is
sufficient to couple to field theories with conserved dipole moment, to better
appreciate the structure of the dipole shift symmetry, it is convenient to pass
to a first-order formulation. Here the spatial metric $h_{\mu\nu}$ is exchanged
for a spatial vielbein $e^a_\mu$ and a local $SO(d)$ symmetry which rotates the
spatial one-forms $e^a_{\mu}$ into each other. The dipole symmetry is then
naturally defined as acting in the tangent bundle.

\subsection{First order formulation of Aristotelian geometries}

The first-order formulation of Aristotelian backgrounds has the clock-form
$n_\mu$ and a spatial vielbein $e^a_\mu$, where $a=1,\ldots,d$ is an index
enumerating this basis of spatial one-forms. There must be a zero linear
combination $v^\mu e^a_\mu = 0$ for some vector field $v^\mu$ normalised as
$v^\mu n_\mu = 1$. We can also define the inverse vielbein $e^\mu_a$ using the
completeness relations $v^\mu n_\nu + e^\mu_a e^a_\nu = \delta^\mu_\nu$ and
$e^\mu_a e^b_\mu = \delta^b_a$. The spatial metric $h_{\mu\nu}$ and cometric
$h^{\mu\nu}$ are related in terms of these as
\begin{equation}
  h_{\mu\nu} = \delta_{ab} e^a_\mu e^b_\nu\,, \qquad
  h^{\mu\nu} = \delta^{ab} e_a^\mu e_b^\nu\,.
\end{equation}
The raising and lowering of $a,b,\ldots$ indices is done using $\delta^{ab}$ and
$\delta_{ab}$.  Note that the spatial metric $h_{\mu\nu}$ has $(d+1)(d+2)/2-1$
independent components, while the spatial vielbein $e^a_\mu$ has $d(d+1)$
independent components. This additional $d(d-1)/2$ components in $e^a_\mu$ arise
from a redundancy under local $SO(d)$ rotations that arises from the
decomposition of $h_{\mu\nu}$ into $e^a_{\mu}$. This local $SO(d)$ symmetry acts
as $e^a_{\mu}\to R^a{}_b e^b_{\mu}$ for $R$ a rotation matrix.

We also introduce a spin connection $\omega^a{}_{b\mu}$ associated with the
$SO(d)$ symmetry,
\begin{equation}
  \omega^a{}_{b\mu}
  = e_\lambda^a \dow_\mu e^\lambda_b + e_\lambda^a \Gamma^\lambda{}_{\mu\nu}
  e^\nu_b,
\end{equation}
where $\Gamma^\lambda{}_{\mu\nu}$ is the connection defined in
Eq. \eqref{eq:connection}. It can be explicitly checked that
$\omega^{ab}{}_{\mu} = -\omega^{ba}{}_{\mu}$ and the associated covariant
derivative, which acts both on spacetime indices $\mu,\nu$ and flat spatial
indices $a,b$, satisfies the properties
\begin{align}
\begin{split}
  \nabla_\mu \delta_{ab}
  &= - \delta_{cb} \omega^c{}_{a\mu}
  - \delta_{ac} \omega^c{}_{b\mu}
  = 0\,, 
  \\
  \nabla_\mu e^\nu_a
  &= \dow_\mu e^\nu_a + \Gamma^\nu{}_{\mu\rho} e^\rho_a
    - e^\nu_b \omega^b{}_{a\mu} = 0\,, 
    \\
  \nabla_\mu e_\nu^a
  &= \dow_\mu e_\nu^a - \Gamma^\rho{}_{\mu\nu} e_\rho^a
    + e_\nu^b \omega^a{}_{b\mu}
    = - \half n_\nu e^{a\rho} \lie_v h_{\mu\rho}\,.
\end{split}
\end{align}
If we were to introduce torsion into the Aristotelian connection
\eqref{eq:connection} (see Footnote \ref{foot:torsion}), it would
correspondingly introduce a torsion into the spin connection.  We can define the
associated spatial torsion and curvature tensors as
\begin{align}
\begin{split}
  T^a{}_{\mu\nu}
  &= \dow_\mu e^a_\nu - \dow_\nu e^a_\mu
    + \omega^a{}_{b\mu} e^b_\nu - \omega^a{}_{b\nu} e^b_\mu\,,
    \\
  R^a{}_{b\mu\nu}
  &= \dow_\mu \omega^{a}{}_{b\nu}
  - \dow_\nu \omega^{a}{}_{b\mu}
  + \omega^a{}_{c\mu}\omega^c{}_{b\nu}
  - \omega^a{}_{c\nu}\omega^c{}_{b\mu}\,.
\end{split}
\end{align}
These are related to the full Aristotelian torsion and curvature tensors as
\begin{align}
\begin{split}
  T^\lambda{}_{\mu\nu}
  &= e^\lambda_a T^a{}_{\mu\nu}
    - 2n_{[\mu}\nabla_{\nu]} v^\lambda
    + v^\lambda F^n_{\mu\nu} \,,
    \\
  R^\lambda{}_{\sigma\mu\nu}
  &= e^\lambda_a e^b_\sigma R^a{}_{b\mu\nu}
    + 2n_\sigma \nabla_{[\mu} \nabla_{\nu]} v^\lambda\,.
\end{split}
\end{align}
For our choice of connection, the Aristotelian torsion is simply
$T^\lambda{}_{\mu\nu} = v^\lambda F^n_{\mu\nu}$ and correspondingly
$T^a{}_{\mu\nu} = e_{a\lambda} n_{[\mu} \nabla_{\nu]}v^\lambda$.


The $U(1)$ connection $A_\mu$ is borrowed directly from the second order
formulation, whereas the spatial dipole gauge field is instead taken to be
$a_{ab}$. The covariant spatial dipole gauge field $a_{\mu\nu}$ is defined as
\begin{equation}
  a_{\mu\nu} = a_{ab} e^a_\mu e^b_\nu,
\end{equation}
and automatically satisfies the constraint $v^\mu a_{\mu\nu} = 0$. We can define
the dipole ``connection'' and ``field strength'' (the analogues of
$A^{\lambda}{}_{\mu}$ and $F^{\lambda}{}_{\mu\nu}$) in the first-order
formulation as
\begin{align}
\begin{split}
  A^a{}_\mu
  &= n_\mu v^\rho F_{\rho\sigma} e^{a\sigma}
  + \half\lb h_{\mu}^\rho F_{\rho\sigma} e^{a\sigma}
    + a^{ab} e_{b\mu} \rb\,,
    \\
  F^{a}{}_{\mu\nu}
  &= \dow_\mu A^a{}_{\nu} - \dow_\nu A^a{}_\mu
  + \omega^a{}_{b\mu} A^b{}_\nu
  - \omega^a{}_{b\nu} A^b{}_\mu
  + n_{[\mu} A^b{}_{\nu]} e^{a\rho}e^\sigma_b \lie_v h_{\rho\sigma}\,.
\end{split}
\end{align}

Now, when coupling a field theory to this first-order formulation of the
background, we impose invariance under diffeomorphisms, local rotations, $U(1)$
gauge transformations, and dipole transformations. At the infinitesimal level,
these act on the background fields in the following way. Let $\chi^{\mu}$ be an
infinitesimal diffeomorphism, $\Omega^a{}_b$ an infinitesimal rotation (with
$\Omega_{(ab)} = 0$), $\Lambda$ a $U(1)$ gauge transformation, and $\psi_a$ an
infinitesimal dipole transformation. Collectively denoting the transformation as
$\hat\scX = (\xi^\mu,\Omega^a{}_b,\Lambda,\psi_a)$, the corresponding generator
$\delta_{\hat{\scX}}$ of the transformation acts on the background fields as
\begin{align}
\begin{split}
  \delta_{\hat\scX} n_\mu
  &= \lie_\chi n_\mu\,,
  \\
  \delta_{\hat\scX} e^a_\mu
  &= \lie_\chi e^a_\mu - \Omega^a{}_b e^b_\mu\,,
  \\
  \delta_{\hat\scX} A_\mu
  &= \lie_\chi A_\mu + \dow_\mu\Lambda + e^a_\mu \psi_a\,,
  \\
  \delta_{\hat\scX} a_{ab}
  &= \lie_\chi a_{ab}
    + \Omega^c{}_a a_{cb}
    + \Omega^c{}_b a_{ac}
    + e^\mu_a\nabla_\mu\psi_b + e^\mu_b\nabla_\mu\psi_a\,.
    \label{eq:sym-variations-first}
\end{split}
\end{align}
So defined, the spin and ``dipole'' connections transform as
\begin{align}
\begin{split}
  \delta_{\hat\scX} \omega^a{}_{b\mu}
  &= \lie_\chi \omega^a{}_{b\mu} + \nabla_\mu \Omega^a{}_b\,,
  \\
  \delta_{\hat\scX} A^a_\mu
  &= \lie_\chi A^a{}_\mu - \Omega^a{}_b A^b{}_\mu
  + \nabla_\mu \psi^a
  + \half n_\mu \psi^b e^{a\rho}e_b^\sigma \lie_v h_{\rho\sigma}\,,
\end{split}
\end{align}
so that the spin connection is indeed a connection under local rotations, and
$A^a{}_\mu$ is nearly a connection under dipole transformations. It can be
checked that these symmetry variations lead to the second-order symmetry
variations given in eq. \eqref{eq:background-variations}.  A lengthy computation
also shows that the symmetry generators form an algebra, with
\begin{equation}
  \label{E:WZ}
  [\delta_{\hat{\scX}},\delta_{\hat{\scX}'}] = \delta_{[\hat{\scX},\hat{\scX}']}\,,
\end{equation}
where the commutator transformation is defined as
\begin{equation}
  \begin{alignedat}{2}
  \label{E:commutatorOfSymmetryParameters}
    \chi^{\mu}_{[\hat{\scX}',\hat{\scX}]}
    &= \delta_{\hat\scX'}\chi^\mu
    &&= \lie_{\chi'}\chi^\mu\,,
    \\
    (\Omega_{[\hat{\scX}',\hat{\scX}]})^a{}_b
    &= \delta_{\hat\scX'}\Omega^a{}_b
    &&= \lie_{\chi'}\Omega^a{}_b - \lie_\chi\Omega'^a{}_b 
    + \Omega^a{}_c \Omega'^c{}_b - \Omega'^a{}_c \Omega^c{}_b\,,
    \\
    \Lambda_{[\hat{\scX}',\hat{\scX}]}
    &= \delta_{\hat\scX'}\Lambda
    &&= \lie_{\chi'}\Lambda - \lie_\chi\Lambda' \,,
    \\
    (\psi_{[\hat{\scX}',\hat{\scX}]})_a
    &=  \delta_{\hat\scX'}\psi_a
    &&= \lie_{\chi'}\psi_a - \lie_{\chi}\psi'_a
    + \psi_b \Omega'^b{}_a - \psi'_b \Omega^b{}_a\,.
  \end{alignedat}
\end{equation}
More details can be found in Appendix \eqref{app:consistency}. Applying the
identity~\eqref{E:WZ} to the generating functional leads to Wess-Zumino
consistency conditions, just as for relativistic field theories.

\subsection{Currents and conservation}

The coupling of conserved currents to background sources follows analogous to
our discussion in Subsection~\ref{S:ward}. Directly plugging in the definition
of second-order sources in terms of the first-order ones into the generating
function variation in Eq.~\eqref{eq:actionVariation}, we can obtain
\begin{equation}
  \delta W
  = \int \df ^{d+1} x \sqrt{\gamma}\lb
  - \epsilon^\mu \delta n_\mu
  + \tau^\mu_{~a} \delta e_\mu^a
  + J^\mu \delta A_\mu
  + J^\mu_{~a}\delta A^a{}_{\mu}
  \rb,
  \label{eq:actionVariation-first}
\end{equation}
where the full momentum current is defined as
$\tau^\mu_{~a} = (v^{\mu} \pi^{\nu} + \tau^{\mu\nu} + \tau^{\mu\nu}_{\text{d}})
e_{a\nu}$, including the contributions from momentum density $\pi^\mu$,
symmetric stress tensor $\tau^{\mu\nu}$, and stress contributions from the
multipole currents $\tau^{\mu\nu}_{\text{d}}= - A^{\mu}_{\rho}J^{\rho\nu} $. The
dipole current is defined as $J^\mu_{~a} = J^{\mu}_{~\lambda} e^\lambda_a$. This
coupling structure, of course, looks more natural from a field theoretic
perspective.

Invariance of the action under $SO(d)$ rotations leads to the relation
$\tau^{\mu}_{[a}e_{b]\mu} = -J^\mu_{~[a}A_{b]\mu}$ that is identically
satisfied. On the other hand, demanding invariance under the remaining symmetry
variations in \eqref{eq:sym-variations-first} lead to the Ward identities in the
first-order form
\begin{align}
\begin{split}
  \nabla'_\mu \epsilon^\mu
  &= - v^\mu f_\mu
  - \tau^{\mu}_{~a} e^a_\nu \nabla_{\mu} v^\nu\,,
  \\
  \nabla'_\mu \tau^\mu_{~a}
  &= e^\mu_a f_\mu
  - n_\mu\tau^\mu_{~b} e^b_\nu e_a^\rho \nabla_\rho v^\nu\,,
  \\
  \nabla'_\mu J^\mu
  &= 0\,,
  \\
  \nabla'_\mu J^{\mu}_{~a}
  &= e_{a\mu} J^\mu\,,
  \label{eq:WardIdentity-first} 
\end{split}
\end{align}
The power-force density $f_\mu$ has already been defined in
Eq.~\eqref{eq:force-density}.

Finally, the dipole transformation properties of various currents can directly
be derived from Eq.~\eqref{eq:dipole-var-currents}. We find
\begin{align}
\begin{split}
  \epsilon^\mu
  &\to \epsilon^\mu
    + \lb 2J^{\mu(\rho}\psi^{\sigma)} - J^{\rho\sigma} \psi^\mu \rb
    \half\lie_v h_{\rho\sigma}\,,
    \\
  \tau^{\mu}_{~a}
  &\to \tau^{\mu}_{~a}
  - J^\mu \psi_a
  + \nabla'_\nu\lb\psi^\nu J^{\mu}_{~a} - J^{\nu}_{~a} \psi^\mu\rb
  + v^\mu J^{\rho}_{~a} F^n_{\rho\sigma} \psi^\sigma\,,
  \label{eq:dipole-var-currents-first}
\end{split}
\end{align}
while $J^\mu$ and $J^\mu_{~a}$ are invariant.

\subsection{Dipole algebra in curved space}
\label{S:dipoleAlgebra}

We have verified the consistency of the symmetry algebra as a Lie algebra. But
one might still wonder what this algebra has to do with the original dipole
symmetry algebra of the flat space theory in Eq.~\eqref{eq:dipole-algebra}. To
make this connection, let us follow the analysis of~\cite{Jain:2018jxj} and
consider a symmetry variation $\delta_{\hat\scX}$, given in terms of the
parameters $\hat\scX = (\chi^\mu,\Omega^a{}_b,\Lambda,\psi_a)$, and formally
decompose it in a basis as
\begin{equation}
  \delta_{\hat\scX} =
  i \chi^\mu n_\mu {\rm H}
  - i\chi^\mu e_\mu^a {\rm P}_a
  + \frac{i}{2} (\Omega^{ab} + \chi^\mu \omega^{ab}{}_{\mu}) {\rm M}_{ab}
  - i (\Lambda + \chi^\mu A_\mu) {\rm Q}
  - i (\psi^a + \chi^\mu A^a{}_\mu) {\rm D}_a.
  \label{eq:dB-decomposition}
\end{equation}
The Hamiltonian ${\rm H}$, momenta ${\rm P}_a$, angular momenta ${\rm M}_{ab}$,
charge ${\rm Q}$, and dipole moment ${\rm D}_a$ should be understood as
generators of local time translations, spatial translations, rotations, $U(1)$
transformations, and dipole transformations in a general curved
background. Inserting this decomposition into the infinite-dimensional symmetry
algebra,~\eqref{E:WZ}, we find that it can be expressed in terms of a finite set
of commutators involving ${\rm H}$, ${\rm P}_a$, etc.:
\begin{align}
\begin{split}
  [{\rm H},{\rm P}_a] &= i v^\mu e_a^\nu {\cal C}_{\mu\nu}\,, 
  \hspace{5.9em}
  [{\rm P}_a,{\rm P}_b] = -i e_a^\mu e_b^\nu {\cal C}_{\mu\nu}\,, \\
  [{\rm H},{\rm D}_a]
  &= - \frac{i}{2} e_a^{\rho} e^{b\sigma} \lie_v h_{\rho\sigma} {\rm D}_b\,,
  \hspace{2em}
  [{\rm P}_a,{\rm D}_b] = i\delta_{ab} {\rm Q}\,,  \\
  [{\rm M}_{ab},{\rm D}_c]
  &= i(\delta_{ac}{\rm D}_b - \delta_{bc}{\rm D}_a)\,, \\
  [{\rm M}_{ab},{\rm P}_c]
  &= i(\delta_{ac}{\rm P}_b - \delta_{bc}{\rm P}_a)\,, \\
  [{\rm M}_{ab},{\rm M}_{cd}] &= i(\delta_{ac}{\rm M}_{bd}
  - \delta_{bc}{\rm M}_{ad} - \delta_{ad}{\rm M}_{bc} + \delta_{bd}{\rm M}_{ac})\,,
  \label{eq:curved-dipole-algebra}
\end{split}
\end{align}
and all other commutators zero. We have defined the ``curvature'' operator
\begin{equation}
  {\cal C}_{\mu\nu}
  = - F^n_{\mu\nu} {\rm H}
  + 2T^a{}_{\mu\nu} {\rm P}_a
  - \half R^{ab}{}_{\mu\nu} {\rm M}_{ab}
  + F^a{}_{\mu\nu}  {\rm D}_a.
  \label{eq:curvature-operator}
\end{equation}
To obtain this we made use of the identities
\begin{align}
\begin{split}
  \delta_{\hat\scX'}(\Omega^{ab} + \chi^\mu \omega^{ab}{}_{\mu})
  &= \lie_{\chi'} (\Omega^{ab} + \chi^\mu \omega^{ab}{}_{\mu})
    + 2\Omega'^{[a}{}_c(\Omega^{b]c} + \chi^\mu \omega^{b]c}{}_{\mu})\,,
    \\
  \delta_{\hat\scX'}(\Lambda + \chi^\mu A_\mu)
  &= \lie_{\chi'}(\Lambda + \chi^\mu A_\mu)
    + \chi^\mu e_\mu^a \psi'_a\,,
    \\
  \delta_{\hat\scX'}(\psi^a + \chi^\mu A_\mu^a)
  &= \lie_{\chi'}(\psi^a + \chi^\mu A_\mu^a)
    - \Omega'^a{}_b(\psi^b + \chi^\mu A_\mu^b)
    \\
  &\qquad
    + (\Omega^a{}_b + \chi^\mu \omega^a{}_{b\mu})\psi'^b
    + \half \chi^\mu n_\mu \psi'^b e^{a\rho}e_b^{\sigma} \lie_v h_{\rho\sigma}\,.
\end{split}
\end{align}
Further details can be found in Appendix \ref{app:algebra}. Note that the $U(1)$
field strength $F_{\mu\nu}$ does not appear in the curvature operator on its own
as it does not transform homogeneously under dipole
transformations.

Eq.~\eqref{eq:curved-dipole-algebra} is the dipole algebra generalised to curved
space. It reduces to the original dipole algebra given in
Eq.~\eqref{eq:dipole-algebra} in the flat space limit. On curved space, the
algebra is almost the same, except that the mutual commutators of ${\rm H}$ and
${\rm P}_a$ are not zero, but are sourced by the curvature operator
${\cal C}_{\mu\nu}$. This should be physically expected because local
translations on curved spacetime do not commute.  $[{\rm H},{\rm D}_a]$ is also
nonzero in curved space. The source of this commutator is proportional to
$\lie_v h_{\mu\nu}$, which can be understood as the ``time derivative'' of the
spatial metric in the reference frame of the background observer with velocity
$v^\mu$. In other words, the dipole moment is no longer conserved when the
system is coupled to a time-dependent spatial metric. This should also be
physically expected because a dipole is a non-local degree of freedom and the
associated dipole moment is sensitive to the spatial separation between the
charges making up the dipole.

\section{Conserved multipole moments}

\label{sec:multipole-moments}

The techniques that we have developed in this work can easily be adapted to
covariantise field theories with higher conserved multipole moments. A system
with $2^n$-pole symmetry has a series of conserved multipole charges
${\rm Q}^{(r)}_{a_1\ldots a_r}$, for $r=0,\ldots, n$, obeying the ``multipole
algebra''~\cite{Gromov:2018nbv}
\begin{align}
\begin{split}
\label{E:multipoleAlgebraFlat}
  [{\rm P}_a,{\rm Q}^{(r)}_{a_1\ldots a_r}]
 & = ir\delta_{a(a_1}{\rm Q}^{(r-1)}_{a_2\ldots a_{r})}\,,
	\\
  [{\rm M}_{ab},{\rm Q}^{(r)}_{c_1\ldots c_r}]
  &= ir\lb \delta_{a(c_1}{\rm Q}^{(r)}_{c_2\ldots c_r)b}
  - \delta_{b(c_1}{\rm Q}^{(r)}_{c_2\ldots c_r)a}\rb\,,
  \\
  [{\rm M}_{ab},{\rm P}_c] &= i(\delta_{ac}{\rm P}_b - \delta_{bc}{\rm P}_a)\,,
  \\
  [{\rm M}_{ab},{\rm M}_{cd}] &= i(\delta_{ac}{\rm M}_{bd}
  - \delta_{bc}{\rm M}_{ad} - \delta_{ad}{\rm M}_{bc} + \delta_{bd}{\rm M}_{ac})\,,
\end{split}
\end{align}
and all other commutators zero. The $U(1)$ monopole charge is
${\rm Q} = {\rm Q}^{(0)}$, while the dipole moment is
${\rm D}_a = {\rm Q}^{(1)}_a$. The $r^{\rm th}$ moment charge (i.e. $2^r$-pole
moment) commutes with the Hamiltonian and transforms as an rank-$r$ tensor under
rotations. Under translations, on the other hand, it picks up the
$(r-1)^{\rm th}$ moment charge. For example, $[P_a,Q]=0$,
$[{\rm P}_a,{\rm D}_{b}] = i\delta_{ab}{\rm Q}$, and
$[{\rm P}_a,{\rm Q}_{bc}^{(2)}] = i(\delta_{ab}{\rm D}_{c} + \delta_{ac}{\rm
  D}_{b})$. Note that the monopole charge commutes with everything, while the
dipole moment obeys the same algebra as noted previously.

In a field theory, $2^n$-pole symmetry charges can be realised by a $U(1)$
charge density $J^t$, obeying the conservation equation of the form
\begin{equation}
  \dow_t J^t + \dow_{i_1}\ldots \dow_{i_{n+1}} J^{i_1\ldots i_{n+1}}_{(n)} = 0\,,
  \label{eq:multi-cons}
\end{equation}
where $J^{i_1\ldots i_{n+1}}_{(n)}$ is the totally-symmetric $n^{\rm th}$ pole current. The
conserved multipole moments are
\begin{equation}
  Q^{i_1\ldots i_r}_{(r)} = \int \df^d x\, J^tx^{i_1}\ldots x^{i_r}\,,  \qquad
  \text{for}\quad r=0,1,\ldots n\,.
\end{equation}
Similar to our analysis of models with conserved dipole moment, it is convenient
to exchange the conservation equation \eqref{eq:multi-cons} for a series of Ward
identities
\begin{equation}
  \dow_{\mu}J^{\mu} = 0\,, \qquad\qquad
  \dow_{i} J_{(r)}^{ij_1\ldots j_{r}} = J_{(r-1)}^{j_1\ldots j_{r}}\,,
  \quad\text{for}\quad r=1,\ldots, n\,.
\end{equation}
where $J_{(0)}^i = J^i$. Each successive $r^{\rm th}$ pole current is given in terms of
the divergence of the $(r+1)^{\rm th}$ pole current, all the way up to $r=n-1$. Expressed in
this language, it is straightforward to couple this theory to background fields.

\subsection{Conserved quadrupole moment}

To not get overwhelmed by the number of multipole charges, let us consider first
the case of a theory with conserved quadrupole moment. An action for such a
model with a single complex quantum field $\Phi$ is given by
\begin{equation}
\label{E:Squadrupole}
  S = \int \df t \df^dx \Big( i \Phi^*\partial_t \Phi +
    {\rm D}_{ijk}(\Phi^*,\Phi^*,\Phi^*){\rm D}_{ijk}(\Phi,\Phi,\Phi) -
    V(\Phi^*\Phi)\Big)\,,
\end{equation}
with
\begin{equation}
  {\rm D}_{ijk}(\Phi,\Phi,\Phi) = \Phi^2\partial_i \partial_j \partial_k \Phi-3\Phi \partial_{(i }\Phi \partial_j \partial_{k)}\Phi +2\partial_i \Phi\partial_j \Phi\partial_k \Phi\,.
\end{equation}
Note that the simplest term with spatial derivatives involves six powers of
$\Phi$.

Coupling to curved space for this theory follows along the same lines as before.
We have the normal Aristotelian background sources: the clock-form $n_\mu$,
spatial metric $h_{\mu\nu}$, and $U(1)$ gauge field $A_\mu$. In addition, we
have a symmetric spatial dipole gauge field $a^{(1)}_{\mu\nu}$ and analogously a
totally-symmetric spatial quadrupole gauge field $a^{(2)}_{\mu\nu\rho}$,
satisfying $v^\mu a^{(1)}_{\mu\nu} = v^\mu a^{(2)}_{\mu\nu\rho} = 0$. Under an
infinitesimal diffeomorphism $\chi^{\mu}$, gauge transformation $\Lambda$,
dipole transformation $\psi^{(1)}_{\mu}$ (obeying $v^{\mu}\psi_{\mu}^{(1)}=0$),
and quadrupole transformation $\psi^{(2)}_{\mu\nu}$ (symmetric, and obeying
$v^{\nu}\psi_{\mu\nu}^{(2)}=0$), which we collectively denote as
$\hat\scX = (\chi^\mu,\Lambda,\psi^{(1)}_\mu,\psi^{(2)}_{\mu\nu})$, the
background and quantum fields transform as
\begin{align}
\begin{split}
	\delta_{\hat{\scX}}\Phi & = \lie_{\chi} \Phi + i \Lambda \Phi\,,
	\\
  \delta_{\hat\scX} A_\mu
  &= \lie_\chi A_\mu + \dow_\mu\Lambda + \psi^{(1)}_\mu\,,
  \\
  \delta_{\hat\scX} a^{(1)}_{\mu\nu}
  &= \lie_\chi a^{(1)}_{\mu\nu}
    + h_{\mu}^\rho h_{\nu}^\sigma
    \lb \nabla_\rho \psi^{(1)}_\sigma + \nabla_\sigma\psi^{(1)}_\rho \rb
    + 2\psi^{(2)}_{\mu\nu}\,,
    \\
  \delta_{\hat\scX} a^{(2)}_{\mu\nu\rho}
  &= \lie_\chi a^{(2)}_{\mu\nu\rho}
    + h_{\mu}^\sigma h_{\nu}^\tau h_{\rho}^\lambda
    \lb \nabla_\sigma \psi^{(2)}_{\tau\lambda}
    + \nabla_\tau\psi^{(2)}_{\lambda\sigma}
    + \nabla_\lambda\psi^{(2)}_{\sigma\tau} \rb\,,
    \label{eq:background-variations-quad}
\end{split}
\end{align}
while $n_\mu$ and $h_{\mu\nu}$ transform as before. The numerical factor of $2$
in the transformation of $a_{\mu\nu}^{(1)}$ is chosen for convenience. The
covariantised version of the quadrupole-symmetric theory~\eqref{E:Squadrupole}
is then invariant under diffeomorphisms, $U(1)$ gauge transformations, dipole
transformations, and quadrupole transformations.

Moving on, we had already defined a dipole ``connection'' and ``field strength''
in Eqs.~\eqref{dipole-connection} and~\eqref{E:dipoleField}, which we reprise
here,
\begin{align}
\begin{split}
  A_{(1)}^\lambda{}_{\mu}
  &= n_\mu v^\rho F_{\rho\sigma} h^{\sigma\lambda}
  + \half\lb h_{\mu}^\rho F_{\rho\sigma} h^{\lambda\sigma}
    + a^{(1)}_{\mu\rho}h^{\rho\lambda} \rb\,, 
    \\
  F_{(1)}^\lambda{}_{\mu\nu}
  &= \nabla_\mu A_{(1)}^\lambda{}_{\nu} - \nabla_\nu A_{(1)}^\lambda{}_{\mu}
  + F^n_{\mu\nu} v^\rho A_{(1)}^\lambda{}_{\rho}
  + 2n_{[\mu} A_{(1)}^\rho{}_{\nu]} \nabla_\rho v^\lambda\,.
\end{split}
\end{align}
These objects transform ``nicely'' under the dipole symmetry, but not under the
quadrupole symmetry, similar to how $A_\mu$ and $F_{\mu\nu}$ transform nicely
under the $U(1)$ monopole symmetry but not under the dipole symmetry. To wit, we
have
\begin{align}
\begin{split}
  \delta_{\hat\scX} A_{(1)}^\lambda{}_{\mu}
  &= \lie_\chi A_{(1)}^\lambda{}_{\mu}
    + \nabla_\mu \psi^\lambda_{(1)}
    + n_\mu \psi^\nu_{(1)} \nabla_\nu v^\lambda
    + \psi^{\lambda\nu}_{(2)} h_{\mu\nu}\,,
    \\
  \delta_{\hat\scX} F_{(1)}^\lambda{}_{\mu\nu}
  &= \lie_\chi F_{(1)}^\lambda{}_{\mu\nu}
    + {\cal R}^\lambda{}_{\rho\mu\nu} \psi^\rho_{(1)} 
    + h^{\lambda\rho}\nabla_\mu \psi^{(2)}_{\nu\rho}
    - h^{\lambda\rho}\nabla_\nu \psi^{(2)}_{\mu\rho}
    + 2n_{[\mu} \psi^\rho_{(2)\nu]} \nabla_\rho v^\lambda\,,
\end{split}
\end{align}
where
\begin{equation}
  {\cal R}^\lambda{}_{\rho\mu\nu}
  = h^\sigma_\rho\lb R^{\lambda}{}_{\sigma\mu\nu} + F^n_{\mu\nu} \nabla_\sigma v^\lambda
  - 2 n_{[\mu}\nabla_{\nu]} \nabla_\sigma v^\lambda \rb.
  \label{eq:calR}
\end{equation}
Similarly, we define the quadrupole ``connection''
\begin{align}
  A_{(2)}^{\lambda\tau}{}_{\mu}
  &= n_\mu v^\rho F^{(\lambda}{}_{(1)\rho\nu}h^{\tau)\nu}
  + \frac{1}{3} \lb 2h_\mu^\rho F^{(\lambda}{}_{(1)\rho\sigma} h^{\tau)\sigma}
    + a^{(2)}_{\mu\rho\sigma} h^{\rho\lambda} h^{\sigma\tau} \rb,
\end{align}
which transforms as 
\begin{equation}
  \delta_{\hat\scX} A_{(2)}^{\lambda\tau}{}_{\mu}
  = \lie_\chi A_{(2)}^{\lambda\tau}{}_{\mu}
    + \lb n_\mu v^\rho + \frac{2}{3} h_\mu^\rho \rb
    {\cal R}^{(\lambda}{}_{\sigma\rho\nu} h^{\tau)\nu} \psi^\sigma_{(1)}  
    + \nabla_\mu\psi_{(2)}^{\lambda\tau} 
    + 2n_\mu\psi_{(2)}^{\sigma(\lambda} \nabla_\sigma v^{\tau)}\,.
\end{equation}
The quadrupole ``field strength,'' on the other hand, is defined as
\begin{align}
\begin{split}
  F_{(2)}^{\lambda\tau}{}_{\mu\nu}
  &= \nabla_{\mu} A_{(2)}^{\lambda\tau}{}_{\nu}
    - \nabla_{\nu} A_{(2)}^{\lambda\tau}{}_{\mu}
    - F^n_{\mu\nu} v^\rho A_{(2)}^{\lambda\tau}{}_{\rho}
    \\
  &\qquad
    + 2A_{(1)}^\sigma{}_{[\mu}  \lb n_{\nu]} v^\rho + \frac{2}{3} h_{\nu]}^\rho \rb
    {\cal R}^{(\lambda}{}_{\sigma\rho\nu} h^{\tau)\nu}
    + 4 n_{[\mu}A_{(2)}^{\rho(\lambda}{}_{\nu]} \nabla_\rho v^{\tau)}\,.
\end{split}
\end{align}


The utility of all this structure is that we can now define symmetry currents as
conjugate to external fields via
\begin{equation}
  \delta W
  = \int\df^{d+1}x \sqrt{\gamma}\lb
  - \epsilon^\mu \delta n_\mu
  + \lb v^\mu \pi^\nu + \half \tau^{\mu\nu} \rb \delta h_{\mu\nu}
  + J^\mu \delta A_\mu
  + J_{(1)}^{\mu}{}_{\lambda}\delta A_{(1)}^\lambda{}_{\mu}
  + J_{(2)}^{\mu}{}_{\lambda\tau}\delta A_{(2)}^{\lambda\tau}{}_{\mu}
  \rb.
  \label{eq:actionVariation-quad}
\end{equation}
Invariance under diffeomorphisms, gauge transformations, etc., leads to Ward
identities as in our analysis in Subsection~\ref{S:ward},
\begin{align}
\begin{split}
  \nabla'_\mu \epsilon^\mu
  &= - v^\mu f_\mu
    - \lb \tau^{\mu\nu}  + \tau_{\text{d}}^{\mu\nu}\rb
    h_{\nu\lambda}\nabla_{\mu} v^{\lambda}\,,
  \\
  \nabla'_\mu\lb v^\mu \pi^\nu + \tau^{\mu\nu} + \tau_{\text{d}}^{\mu\nu} \rb
  &= h^{\nu\mu} f_\mu
    - \pi_\mu h^{\nu\lambda} \nabla_\lambda v^\mu\,,
    \\
  \nabla'_\mu J^\mu
  &= 0\,, 
  \\
  \nabla'_\mu J_{(1)}^{\mu\nu}
  &= h^{\nu}_\mu J^\mu\,,
  \\
  \nabla'_\mu J_{(2)}^{\mu\nu\rho}
  &= J_{(1)}^{\nu\rho}\,.
\end{split}
\end{align}
The derivatives on the left-hand-side are as in Subsection~\ref{S:ward},
$\nabla'_{\mu} =\nabla_{\mu}+F^n_{\mu\nu}v^{\nu}$.  The power-force density gets
modified to
\begin{align}
\begin{split}
  f_\mu
  &= -F^n_{\mu\nu} \epsilon^\nu
    - h_{\mu\lambda} A^\lambda{}_{\nu} J^\nu
    + J_{(1)}^{\nu}{}_{\lambda}
    \lb F_{(1)}^{\lambda}{}_{\mu\nu} - h_{\nu\rho}A_{(2)}^{\rho\lambda}{}_{\mu} \rb
    + F_{(2)}^{\lambda\tau}{}_{\mu\nu} J_{(2)}^{\nu}{}_{\lambda\tau}
    \\
  &\qquad\qquad 
    - n_\mu \lb A_{(1)}^{\rho}{}_{\nu} J_{(1)}^{\nu}{}_{\lambda} 
    + 2 A_{(2)}^{\rho\tau}{}_{\nu} J_{(2)}^{\nu}{}_{\lambda\tau} \rb \nabla_{\rho}
    v^\lambda 
    \\
  &\qquad\qquad \qquad 
    - \lb n_\mu v^\sigma 
    A_{(1)}^{\rho}{}_{\nu}
    + 2 
    h_{[\mu}^\sigma A_{(1)}^{\rho}{}_{\nu]} \rb J_{(2)}^{\nu}{}_{\lambda\tau}
    {\cal R}^{\lambda}{}_{\rho\sigma\beta} h^{\tau\beta}\,,
\end{split}
\end{align}
whereas $\tau_{\text{d}}^{\mu\nu}$ now has contributions from the quadrupole
current as well
\begin{equation}
  \tau_{\text{d}}^{\mu\nu}
  = - A_{(1)}^{\mu}{}_{\rho} J^{\rho\nu}_{(1)} 
  - 2 A_{(2)}^{\mu\lambda}{}_{\rho} J_{(2)}^{\rho\nu}{}_{\lambda}.
\end{equation}

\subsection{Conserved higher moments}

This construction can easily be generalised to systems with higher multipole
symmetry. For a system with conserved $n^{\rm th}$ moments, the background fields
comprise of the clock form $n_\mu$, spatial metric $h_{\mu\nu}$, gauge field
$A_\mu$, and a series of spatial multipole gauge fields
$a^{(r)}_{\mu_1\ldots\mu_{r+1}}$ for $r=1,\ldots, n$. Their transformation laws
are 
\begin{align}
  \delta_{\hat\scX} A_\mu
  &= \lie_\chi A_\mu + \dow_\mu\Lambda + \psi^{(1)}_\mu\,, \nn\\
  \delta_{\hat\scX} a^{(r)}_{\mu_1\ldots\mu_{r+1}}
  &= \lie_\chi a^{(r)}_{\mu_1\ldots\mu_{r+1}}
    + (r+1)\, h_{\mu_1}^{\rho_1} \ldots h_{\mu_{r+1}}^{\rho_{r+1}}
    \nabla_{(\rho_1} \psi^{(r)}_{\rho_2\ldots\rho_{r+1})}
    + (r+1)\,\psi^{(r+1)}_{\mu_1\ldots\mu_{r+1}} \,,\qquad
    \text{for}~ r<n\,, \nn\\
  \delta_{\hat\scX} a^{(n)}_{\mu_1\ldots\mu_{n+1}}
  &= \lie_\chi a^{(n)}_{\mu_1\ldots\mu_{n+1}}
    + (n+1)\, h_{\mu_1}^{\rho_1} \ldots h_{\mu_{n+1}}^{\rho_{n+1}}
    \nabla_{(\rho_1} \psi^{(n)}_{\rho_2\ldots\rho_{n+1})}\,,
    \label{eq:background-variations-higher}
\end{align}
while $n_\mu$ and $h_{\mu\nu}$ transform as usual. So defined, the symmetry generators obey an algebra
\begin{equation}
  [\delta_{\hat{\scX}},\delta_{\hat{\scX}'}] = \delta_{[\hat{\scX},\hat{\scX}']}\,,
\end{equation}
with a suitably defined commutator, so that one can speak of Wess-Zumino
consistency conditions for the currents of these theories. Using these
variations, we can define multipole ``connections'' and ``field strengths'' as
\begin{align}
	\nn
  A_{(r)}^{\lambda_1\ldots \lambda_{r}}{}_{\mu}
  &= n_\mu v^\rho
    F_{(r-1)}^{(\lambda_1\ldots \lambda_{r-1}}{}_{\rho\nu}h^{\lambda_{r})\nu}
    + \frac{1}{r+1}
    \lb r h^\rho_\mu
    F_{(r-1)}^{(\lambda_1\ldots\lambda_{r-1}}{}_{\rho\nu}h^{\lambda_{r})\nu}
    + a_{(r)}^{\lambda_1\ldots\lambda_{r}}{}_{\mu} \rb\,,
    \\
  F_{(r)}^{\lambda_1\ldots \lambda_{r}}{}_{\mu\nu} 
  &= 2\nabla_{[\mu} A_{(r)}^{\lambda_1\ldots \lambda_{r}}{}_{\nu]}
    + v^{\rho} F^n_{\mu\nu} A_{(r)}^{\lambda_1\ldots\lambda_r}{}_{\rho}
    - 2\sum_{s=1}^{r} A_{(s)}^{\rho_1\ldots\rho_s}{}_{[\mu}
    {\cal Y}^{(r,s)}_{\nu]}{}^{\lambda_1\ldots
    \lambda_{r}}_{\rho_1\ldots\rho_s}\,.
    \label{eq:multipole-connection}
\end{align}
Here
${\cal Y}^{(r,s)}_{\mu}{}^{\lambda_1\ldots \lambda_{r}}_{\rho_1\ldots\rho_s}$
are some $(r+s+1)$-rank tensors that are entirely given in terms of the
curvature $R^\lambda{}_{\rho\mu\nu}$, frame acceleration $\nabla_\mu v^\nu$, and
their derivatives. See Appendix \ref{app:multipole-connection} for more details.
The ``connections'' transform as 
\begin{align}
  \delta_{\hat{\scX}} A_{(r)}^{\lambda_1\ldots \lambda_{r}}{}_{\mu}
  &= \lie_\chi A_{(r)}^{\lambda_1\ldots \lambda_{r}}{}_{\mu}
    + \nabla_\mu \psi^{\lambda_1\ldots \lambda_r}_{(r)}
    + \sum_{s=1}^{r}
    {\cal Y}_\mu^{(r,s)}{}^{\lambda_1\ldots\lambda_{r}}_{\nu_1\ldots \nu_s}
    \psi_{(s)}^{\nu_1\ldots \nu_s} 
    + h_{\mu\nu} \psi^{\nu\lambda_1\ldots\lambda_r}_{(r+1)}\,,
    \label{eq:multipole-connection-var}
\end{align} 
and the variation of the ``field strengths'' is given in the Appendix.  The
symmetry currents are then defined by the variation
\begin{align}
  \delta W
  &= \int\df^{d+1}x \sqrt{\gamma} \bigg[
  - \epsilon^\mu \delta n_\mu
  + \lb v^\mu \pi^\nu + \half \tau^{\mu\nu} \rb \delta h_{\mu\nu}
  + j^\mu \delta A_\mu 
    + \sum_{r=1}^n J_{(r)}^{\mu}{}_{\lambda_1\ldots\lambda_{r}}
    \delta A_{(r)}^{\lambda_1\ldots\lambda_r}{}_{\mu}
  \bigg]\,.
  \label{eq:actionVariation-multipole}
\end{align}
The Ward identities read
\begin{align}
\begin{split}
  \nabla'_\mu \epsilon^\mu
  &= - v^\mu f_\mu
    - \lb \tau^{\mu\nu} + \tau_{\text{d}}^{\mu\nu}\rb
    h_{\nu\lambda}\nabla_{\mu} v^{\lambda}\,,
  \\
  \nabla'_\mu\lb v^\mu \pi^\nu + \tau^{\mu\nu} + \tau_{\text{d}}^{\mu\nu} \rb
  &= h^{\nu\mu} f_\mu
    - \pi_\mu h^{\nu\lambda} \nabla_\lambda v^\mu\,,
    \\
    \nabla'_\mu J^\mu
  &= 0\,,
  \\
  \nabla'_\mu J^{\mu\nu_1\ldots\nu_r}_{(r)}
  &= J^{\nu_1\ldots\nu_r}_{(r-1)}\,,
\end{split}
\end{align}
where the power-force density gets modified to
\begin{align}
  f_\mu
  &= -F^n_{\mu\nu} \epsilon^\nu
    - h_{\mu\lambda} A_{(1)}^\lambda{}_{\nu} J^\nu \nn\\
  &\qquad
    + \sum_{r=1}^{n-1} 
    \lb F_{(r)}^{\lambda_1\ldots\lambda_r}{}_{\mu\nu} 
    - h_{\nu\rho}
    A_{(r+1)}^{\rho\lambda_1\ldots\lambda_{r}}{}_{\mu} \rb
    J_{(r)\lambda_1\ldots\lambda_{r}}{}^{\nu}
    + F_{(n)}^{\lambda_1\ldots\lambda_n}{}_{\mu\nu}
    J_{(n)\lambda_1\ldots\lambda_{n}}{}^{\nu} \nn
    \\
  &\qquad
    - \sum_{r=1}^n \sum_{s=1}^{r}
   \lb {\cal Y}^{(r,s)}_{\mu}{}^{\lambda_1\ldots
    \lambda_{r}}_{\rho_1\ldots\rho_s}
    A_{(s)}^{\rho_1\ldots\rho_s}{}_{\nu}
    - {\cal Y}^{(r,s)}_{\nu}{}^{\lambda_1\ldots
    \lambda_{r}}_{\rho_1\ldots\rho_s}
    A_{(s)}^{\rho_1\ldots\rho_s}{}_{\mu}\rb
    J_{(r)\lambda_1\ldots\lambda_{r}}{}^{\nu},
\end{align}
while $\tau_{\text{d}}^{\mu\nu}$ gets contributions from all the multipole currents
\begin{equation}
  \tau_{\text{d}}^{\mu\nu}
  = - \sum_{r=1}^n r h^{\nu\lambda_r}
  A_{(r)}^{\mu\lambda_1\ldots\lambda_{r-1}}{}_{\rho}
  J_{(r)\lambda_1\ldots\lambda_{r}}{}^{\rho}\,.
\end{equation}

\subsection{Multipole algebras in curved space}

As in our discussion of the dipole algebra in curved space, we can understand
the commutator of symmetry generators
$[\delta_{\hat{\scX}},\delta_{\hat{\scX}'}] = \delta_{[\hat{\scX},\hat{\scX}']}$
in terms of commutators for a finite set of operators, giving the curved space
version of the multipole algebra~\cite{Jain:2018jxj}. To do this efficiently, we
need to pass to the first-order formulation discussed in Section
\ref{sec:first-order}. Without going into even more detail, we note that the
first order versions of multipole ``connections,'' ``field strengths,'' and
symmetry parameters are defined naturally as
$A_{(r)}^{a_1\ldots a_r}{}_{\mu} = e^{a_1}_{\lambda_1}\ldots
e^{a_r}_{\lambda_r}A_{(r)}^{\lambda_1\ldots\lambda_r}{}_{\mu}$,
$F_{(r)}^{a_1\ldots a_r}{}_{\mu\nu} = e^{a_1}_{\lambda_1}\ldots
e^{a_r}_{\lambda_r}F_{(r)}^{\lambda_1\ldots\lambda_r}{}_{\mu\nu}$, and
$\psi^{a_1\ldots a_r}_{(r)} = e^{a_1}_{\lambda_1}\ldots
e^{a_r}_{\lambda_r}\psi^{\lambda_1\ldots\lambda_r}_{(r)}$ respectively. A
symmetry transformation parametrised by
$\hat\scX=(\chi^\mu,\Omega^a{}_b,\Lambda,\psi^{a_1\ldots a_r}_{(r)})$ can be
decomposed into a basis of generators according to
\begin{align}
\begin{split}
  \delta_{\hat\scX}
  &=
  i \chi^\mu n_\mu {\rm H}
  - i\chi^\mu e_\mu^a {\rm P}_a
  + \frac{i}{2} (\Omega^{ab} + \chi^\mu \omega^{ab}{}_{\mu}) {\rm M}_{ab}
    - i (\Lambda + \chi^\mu A_\mu) {\rm Q} 
    \\
  &\qquad
    - i \sum_{r=1}^n \frac{1}{r!} \lb \psi_{(r)}^{a_1\ldots a_{r}}
    + \chi^\mu A_{(r)}^{a_1\ldots a_{r}}{}_\mu \rb {\rm Q}_{a_1\ldots a_{r}}^{(r)}\,.
\end{split}
\end{align}
Plugging this parameterisation into the infinite-dimensional algebra
$[\delta_{\hat{\scX}},\delta_{\hat{\scX}'}]=\delta_{[\hat{\scX},\hat{\scX}']}$,
we find that the infinite-dimensional algebra is equivalent to a finite set of
commutators, in the form of the curved space algebra
\begin{equation}
\begin{alignedat}{2}
  [{\rm H},{\rm P}_a]
  &= i v^\mu e_a^\nu {\cal C}_{\mu\nu}\,,
  \hspace{6.8em}
  [{\rm P}_a,{\rm P}_b]
  = -i e_a^\mu e_b^\nu {\cal C}_{\mu\nu}\,, \\
  [{\rm H},{\rm Q}^{(r)}_{a_1\ldots a_r}]
  &= -i v^\mu {\cal Y}^{(r)}_{\mu a_1\ldots a_r}\,,
  \hspace{3em}
  [{\rm P}_a,{\rm Q}^{(r)}_{a_1\ldots a_r}]
  = ir\delta_{a(a_1}{\rm Q}^{(r-1)}_{a_2\ldots a_{r})}
  + ie_a^\mu {\cal Y}^{(r)}_{\mu a_1\ldots a_r}\,, \\
  [{\rm M}_{ab},{\rm Q}^{(r)}_{c_1\ldots c_r}]
  &= ir\lb \delta_{a(c_1}{\rm Q}^{(r)}_{c_2\ldots c_r)b}
  - \delta_{b(c_1}{\rm Q}^{(r)}_{c_2\ldots c_r)a}\rb\,, \\
  [{\rm M}_{ab},{\rm P}_c]
  &= i(\delta_{ac}{\rm P}_b - \delta_{bc}{\rm P}_a)\,, \\
  [{\rm M}_{ab},{\rm M}_{cd}] &= i(\delta_{ac}{\rm M}_{bd}
  - \delta_{bc}{\rm M}_{ad} - \delta_{ad}{\rm M}_{bc} + \delta_{bd}{\rm M}_{ac})\,,
\end{alignedat}
\label{E:multipoleAlgebraCurved}
\end{equation}
and other commutators zero. The operator ${\rm Q}^{(0)}$ should be understood as
${\rm Q}$. We have defined the multipole shift parameter
\begin{align}
  {\cal Y}^{(r)}_{\mu a_1\ldots a_r} = \sum_{s=r}^n \frac{r!}{s!}
  v^\mu {\cal Y}^{(s,r)}_\mu{}^{b_1\ldots b_s}_{a_1\ldots a_r}
  {\rm Q}^{(s)}_{b_1\ldots b_s}\,,
\end{align}
while the definition of the curvature operator is also modified to
\begin{align}
  {\cal C}_{\mu\nu}
  &= - F^n_{\mu\nu} {\rm H}
  + 2T^a{}_{\mu\nu} {\rm P}_a
    - \half R^{ab}{}_{\mu\nu} {\rm M}_{ab} 
    + \sum_{r=1}^n \frac{1}{r!}
    \lb F_{(r)}^{a_1\ldots a_r}{}_{\mu\nu}
    + 2e_{b[\mu}A_{(r+1)}^{ba_1\ldots a_r}{}_{\nu]} \rb
    {\rm Q}^{(r)}_{a_1\ldots a_r}.
\end{align}
Along the way we have used
\begin{align}
	\nn
  \delta_{\hat{\scX}'}\lb\psi_{(r)}^{a_1\ldots a_{r}}
  + \chi^\mu A_{(r)}^{a_1\ldots a_{r}}{}_{\mu}\rb
  &= \lie_{\chi'}\lb\psi_{(r)}^{a_1\ldots a_{r}}
  + \chi^\mu A_{(r)}^{a_1\ldots a_{r}}{}_{\mu}\rb
    - r\Omega'^{(a_1}{}_{b} \lb \psi_{(r)}^{a_2\ldots a_{r})b}
    + \chi^\mu A_{(r)}^{a_2\ldots a_{r})b}{}_{\mu}\rb 
    \\
  &\qquad\qquad 
    + r\lb\Omega^{(a_1}{}_{b} + \chi^\mu\omega^{(a_1}{}_{b\mu} \rb
    \psi'^{a_2\ldots a_{r})b}_{(r)}\\
    \nn
  &\qquad \qquad \qquad
    + \sum_{s=1}^{r} \chi^\mu
    {\cal Y}_\mu^{(r,s)}{}^{a_1\ldots a_{r}}_{b_1\ldots b_{s}}
    \psi'^{b_1\ldots b_{s}}_{(s)}
    + \chi^\mu e_{b\mu} \psi'^{a_{1}\ldots a_{r}b}_{(r+1)}\,.
\end{align}

The curved space algebra~\eqref{E:multipoleAlgebraCurved} should be compared
with the flat space algebra~\eqref{E:multipoleAlgebraFlat}. The two agree in
flat space, where the generalised curvatures $\mathcal{C}_{\mu\nu}$ and
$\mathcal{Y}$'s vanish. As in the dipole algebra, $[{\rm H},{\rm P}_a]$ and
$[{\rm P}_a,{\rm P}_b]$ are nonzero in curved space.  Also like before
$[{\rm H},{\rm Q}^{(r)}_{a_1\ldots a_r}]$ is nonzero, affirming that multipole
moments are not time-independent when coupled to generic time-dependent
backgrounds. A novel feature is that $[{\rm P}_a,{\rm Q}^{(r)}_{a_1\ldots a_r}]$
gets contributions from higher moments ${\rm Q}^{p\geq r}_{a_1\ldots a_p}$ and
not just the preceding moment ${\rm Q}^{(r-1)}_{a_1\ldots a_{r-1}}$.

\section{Discussion}

\label{S:discussion}

In this work we have proposed a coupling of simple continuum field theories with
fracton order to curved spacetime backgrounds. The models we considered are, in
flat space, invariant under spacetime translations and rotations, with a
conserved $U(1)$ charge and a conserved dipole moment. A prototypical example of
such a system is the theory of charged matter in~\cite{Pretko:2018jbi}. The
Noether currents associated with these symmetries naturally couple to (i) a
spacetime geometry consisting of a spatial metric $h_{\mu\nu}$ and a clock
one-form $n_{\mu}$, (ii) a $U(1)$ gauge field $A_{\mu}$, and (iii) a source
$a_{ij}$ for the dipole current. The conservation of $U(1)$ charge and dipole
moment in flat space is encoded in two Ward identities, namely
\begin{equation}
  \partial_{\mu}J^{\mu} = 0\,, \qquad J^i = \partial_j J^{ij}\,,
\end{equation}
which when put together yield
$\partial_t J^t + \partial_i \partial_j J^{ij} = 0$. The first is the standard
Ward identity provided that we couple matter fields to a background gauge field
$A_{\mu}$ in such a way that the total action is invariant under local $U(1)$
transformations. The second is enforced by introducing a local ``dipole
symmetry,'' which only acts on the background fields as
\begin{equation}
  \label{E:dipoleShift}
  A_i \to A_i + \psi_i \,, \qquad
  a_{ij} \to a_{ij}+ \partial_i \psi_j + \partial_j \psi_i\,.
\end{equation}
The dipole symmetry could be potentially gauge-fixed by setting $A_i=0$, which
ties the dipole shift parameter to the gauge parameter as
$\psi_i = - \dow_i\Lambda$.  By introducing the background spacetime $n_{\mu}$,
$h_{\mu\nu}$, one can covariantise the theory of charged matter
in~\cite{Pretko:2018jbi}, and generalise the dipole transformation above to
curved space. One can also use the background coupling to derive the relevant
Ward identities associated with energy and momentum conservation in these
systems.

The procedure of coupling to curved background sources follows as usual. Given a
theory with fracton order in flat space, we strive to introduce the background
sources in such a way so as to make the full theory, including background and
dynamical fields, invariant under spacetime diffeomorphisms, local $U(1)$
transformations, and local dipole transformations. At the infinitesimal level,
these symmetry transformations are enacted by first-order linear differential
operators, $\delta_{\hat{\scX}}$, where $\hat{\scX}$ collectively labels the
infinitesimal diffeomorphism, gauge transformation, and dipole
transformation. We showed that $\delta_{\hat{\scX}}$ obey an
infinite-dimensional Lie algebra,\footnote{It turns out that this
  infinite-dimensional algebra is highly redundant. See
  Subsection~\ref{S:dipoleAlgebra} for more detail.}
$[\delta_{\hat{\scX}},\delta_{\hat{\scX}'}] =
\delta_{[\hat{\scX},\hat{\scX}']}$, leading to Wess-Zumino consistency
conditions. In the absence of anomalies, radiative corrections are expected to
preserve these symmetries, leading to Ward identities for correlation
functions. We computed these along with the transformation laws of the various
symmetry currents under dipole transformations.

Along the way, we considered the symmetric tensor gauge theory
of~\cite{Pretko:2016kxt}. This theory of dynamical gauge fields $A_t$, $a_{ij}$
is an analogue of electromagnetism for systems with local dipole-invariance and
naturally couples to the scalar matter theory mentioned above.
In flat space, one can define the analogues of gauge-invariant ``electric'' and
``magnetic'' fields built from $A_t$, $a_{ij}$. However, we found that there is
no gauge-invariant notion of an electric or magnetic field in a general curved
background, suggesting that a covariant curved space theory does not
exist.\footnote{Gauge-invariant combinations do exist on spacetimes with
  absolute time and a time-independent Einstein spatial metric. That result
  matches the previous work of~\cite{Gromov:2017vir,Slagle:2018kqf}.} Indeed, there is an
obstruction to placing this theory in curved space while preserving diffeomorphism invariance, visible already in the flat
space theory. Namely, there is no gauge-invariant definition of the flat space
momentum density and spatial stress tensor. Consequently, the linearized coupling of the flat space theory
to metric perturbations breaks the gauge symmetry, rendering the model
inconsistent. Interestingly, we did find that the Higgs phase of the gauge
theory can be coupled to curved space in a gauge-invariant way.

In the Introduction, we mentioned that perhaps the most experimentally relevant
spacetime symmetry of a fracton model is where $U(1)$ charge, dipole moment, and
the trace of the quadrupole moment are all conserved. In such a theory isolated
charges are immobile, while dipoles can move in a direction perpendicular to
their dipole moment. While we have not discussed this symmetry pattern in the
main text, it is easy to generalise our formalism to these systems. Namely, one
takes the background fields and symmetries we considered in
Section~\ref{S:curved}, but tunes the couplings of the field theory in question
so that the spatial trace of the dipole source $a_{\mu\nu}$ does not appear in
the action, i.e. the spatial trace of the dipole current vanishes,
$h_{\mu\nu}J^{\mu\nu} = 0$. It would be interesting to visit these theories in
greater detail in the future.

So far, several anomalies involving background tensor gauge fields have been
discovered for field theories of fracton order~\cite{Gorantla:2021bda,
  Burnell:2021reh, Yamaguchi:2021xeq}. In the relativistic setting, we are of
course familiar with gravitational and mixed gravitational-flavor anomalies. One
might wonder if there are gravitational anomalies, pure or mixed, in the
landscape of field theories of fractons. In this work, we found evidence that this is the case for the symmetric tensor gauge theory. We found an the obstruction to coupling
the symmetric tensor gauge theory to curved background in a covariant way, which strongly suggests the existence of a mixed gauge-gravitational anomaly
signaling the breakdown of gauge symmetry in curved spacetime. It would
be interesting to uncover if this particular non-invariance can be cured by sacrificing covariance, or by the inflow mechanism, and the role of gravitational anomalies in this landscape
more generally.

Existing field theory models with conserved dipole moment have the feature that
dipoles are composite objects. Dipole moments can either be generated by
radiative corrections in the presence of an elementary charge, or through a
bound state of a charge-anticharge pair. Indeed, one can envision a low-energy
limit of a model where there is an energy gap to the production of charges, and
the lowest-energy degrees of freedom are electrically neutral dipoles. What
would such a field theory look like? A plausible hint comes from our
construction of the ``dipole connection'' $A^\lambda_{~\mu}$, which allows for
coupling to field theories where degrees of freedom can carry an ``intrinsic
dipole moment,'' much like an independent spin connection $\omega^a{}_{b\mu}$
allows for coupling to field theories with intrinsic spin. It will be
interesting to pursue this line of thought further.

Finally, thanks to this work, we have developed an appropriate notion of curved
spacetime for theories with conserved dipole moment, along with the right curved
space symmetries, Ward identities, and the transformation laws for the symmetry
currents. With this information at hand, we are nearly ready to construct the
hydrodynamic effective description of these models. The last ingredient we
require is the low-energy symmetry breaking pattern of interacting fracton
models, which, at least in certain soluble large $N$ models, will be presented
in~\cite{LargeN}.

\subsection*{Acknowledgements}

We would like to thank P.~Glorioso, A.~Gromov, and A.~Raz for enlightening
discussions. AJ would like to thank the University of Victoria, Canada where
part of this project was done. AJ is funded by the European Union’s Horizon 2020
research and innovation programme under the Marie Sk{\l}odowska-Curie grant
agreement NonEqbSK No. 101027527. AJ is also supported by the Netherlands
Organization for Scientific Research (NWO) and by the Dutch Institute for
Emergent Phenomena (DIEP) cluster at the University of Amsterdam.  This work is
supported in part by the NSERC Discovery Grant program of Canada.

\appendix

\section{Calculational details}

\label{app:calc}

In this Appendix we compile some of the calculational details that have been
used in this work.

\subsection{Consistency of symmetry algebra}
\label{app:consistency}

The main goal of this Appendix is to demonstrate that the generators
$\delta_{\hat{\scX}}$ of an infinitesimal symmetry transformation furnish a Lie
algebra.  Doing so requires that we show that the commutator closes, i.e.  the
commutator of two symmetry variations is itself be a symmetry variation,
$[\delta_{\hat\scX},\delta_{\hat\scX'}] = \delta_{[\hat\scX,\hat\scX']}$, for
some appropriately defined commutator operation $[\hat\scX,\hat\scX']$, and that
the variations $\delta_{\hat{\scX}}$ obey the Jacobi identity. Because the
symmetry variations $\delta_{\hat{\scX}}$ are represented by first-order linear
differential operators, they automatically satisfy the Jacobi identity, so we
need only show that the commutator closes.

In fact, we will find that not only do the $\delta_{\hat{\scX}}$ form an
algebra, but so do the infinitesimal transformations $\hat{\scX}$ themselves.
We begin by reprising the variation of the symmetry parameters
$\hat\scX = (\chi^\mu,\Omega^a{}_b,\Lambda,\psi_a)$ given by
Eq.~\eqref{E:commutatorOfSymmetryParameters} as
\begin{equation}
  \begin{alignedat}{2}
    \chi^{\mu}_{[\hat{\scX}',\hat{\scX}]}
    &= \delta_{\hat\scX'}\chi^\mu
    &&= \lie_{\chi'}\chi^\mu\,,
    \\
    (\Omega_{[\hat{\scX}',\hat{\scX}]})^a{}_b
    &= \delta_{\hat\scX'}\Omega^a{}_b
    &&= \lie_{\chi'}\Omega^a{}_b - \lie_\chi\Omega'^a{}_b 
    + \Omega^a{}_c \Omega'^c{}_b - \Omega'^a{}_c \Omega^c{}_b\,,
    \\
    \Lambda_{[\hat{\scX}',\hat{\scX}]}
    &= \delta_{\hat\scX'}\Lambda
    &&= \lie_{\chi'}\Lambda - \lie_\chi\Lambda' \,,
    \\
    (\psi_{[\hat{\scX}',\hat{\scX}]})_a
    &=  \delta_{\hat\scX'}\psi_a
    &&= \lie_{\chi'}\psi_a - \lie_{\chi}\psi'_a
    + \psi_b \Omega'^b{}_a - \psi'_b \Omega^b{}_a\,.
  \end{alignedat}
\end{equation}
Let us compute the action of successive variations on the symmetry parameters
themselves
\begin{align}
\begin{split}
  \delta_{\hat\scX''}\delta_{\hat\scX'}\chi^\mu
  &= \lie_{\chi''}\lie_{\chi'}\chi^\mu \,,
  \\
  \delta_{\hat\scX''}\delta_{\hat\scX'}\Omega^a{}_b
  &= \lie_{\chi''}\lie_{\chi'}\Omega^a{}_b
    + \lie_\chi \lb \lie_{\chi'}\Omega''^a{}_b 
    - \Omega'^a{}_c \Omega''^c{}_b \rb 
    \\
  &\qquad
    - \Omega^a{}_c \lb \lie_{\chi'} \Omega''^c{}_b
    - \Omega'^c{}_d \Omega''^d{}_b \rb
    + \lb \lie_{\chi'} \Omega''^a{}_c 
    + \Omega''^a{}_d \Omega'^d{}_c \rb \Omega^c{}_b 
    \\
  &\qquad
    \textcolor{gray}{
    +\, \lb \Omega''^a{}_c \lie_\chi \Omega'^c{}_b
    + \Omega'^a{}_c \lie_{\chi} \Omega''^c{}_b \rb
    - \lb \Omega''^a{}_d \Omega'^c{}_b 
    + \Omega'^a{}_d \Omega''^c{}_b \rb \Omega^d{}_c} 
    \\
  &\qquad
    \textcolor{gray}{
    +\, \lie_{\chi'}\lb - \lie_\chi\Omega''^a{}_b 
    + \Omega^a{}_c \Omega''^c{}_b - \Omega''^a{}_c \Omega^c{}_b \rb} 
    \\
  &\qquad
    \textcolor{gray}{
    +\, \lie_{\chi''}\lb - \lie_\chi\Omega'^a{}_b 
    + \Omega^a{}_c \Omega'^c{}_b - \Omega'^a{}_c \Omega^c{}_b \rb}\,,
    \\
  \delta_{\hat\scX''}\delta_{\hat\scX'}\Lambda
  &= \lie_{\chi''}\lie_{\chi'}\Lambda
    + \lie_{\chi}\lie_{\chi'}\Lambda'' 
    \\
  &\qquad
    \textcolor{gray}{
    -\, \lie_{\chi'}\lie_\chi\Lambda''
    - \lie_{\chi''}\lie_{\chi}\Lambda'}\,,
    \\
  \delta_{\hat\scX''}\delta_{\hat\scX'}\psi_a
  &= \lie_{\chi''}\lie_{\chi'}\psi_a
    + \lie_{\chi} \lb \lie_{\chi'}\psi''_a + \psi''_b \Omega'^b{}_a \rb \,,
    \\
  &\qquad
    - \psi_b \lb \lie_{\chi'} \Omega''^b{}_a
    - \Omega'^b{}_c \Omega''^c{}_a \rb
    + \lb \lie_{\chi'} \psi''_b
    + \psi''_c \Omega'^c{}_b \rb \Omega^b{}_a \,,
    \\
  &\qquad
    \textcolor{gray}{
    -\, \lb \psi'_b \Omega''^c{}_a
    + \psi''_b \Omega'^c{}_a \rb \Omega^b{}_c
    - \lb \lie_{\chi} \psi'_b \Omega''^b{}_a
    + \lie_\chi \psi''_b \Omega'^b{}_a
    \rb} \,,
    \\
  &\qquad
    \textcolor{gray}{
    +\, \lie_{\chi'} \lb - \lie_{\chi}\psi''_a
    + \psi_b \Omega''^b{}_a - \psi''_b \Omega^b{}_a \rb
    + \lie_{\chi''} \lb - \lie_{\chi}\psi'_a
    + \psi_b \Omega'^b{}_a - \psi'_b \Omega^b{}_a \rb}\,.
\end{split}
\end{align}
The gray terms are symmetric under the exchange of
$\hat\scX'\leftrightarrow\hat\scX''$ and drop out of the commutator. Inspecting
the remaining terms, it is trivial to verify that
$[\delta_{\hat\scX''},\delta_{\hat\scX'}]\hat\scX =
\delta_{[\hat\scX'',\hat\scX']}\hat\scX$. It is also easy to prove the Jacobi
identity from here
\begin{align}
\begin{split}
  [\hat\scX'',[\hat\scX',\hat\scX]]
  = - \delta_{[\hat\scX',\hat\scX]}\hat\scX''
  &= - \delta_{\hat\scX'}\delta_{\hat\scX}\hat\scX''
    + \delta_{\hat\scX}\delta_{\hat\scX'}\hat\scX'' 
    \\
  &= - \delta_{\hat\scX'}[\hat\scX,\hat\scX'']
    - \delta_{\hat\scX}[\hat\scX'',\hat\scX'] 
    \\
  &= - [\hat\scX',[\hat\scX,\hat\scX'']]
    - [\hat\scX,[\hat\scX'',\hat\scX']]\,.
\end{split}    
\end{align}
This demonstrates that the $\hat{\scX}$'s form an infinite-dimensional Lie
algebra.

We now compute the second variation of background fields, finding
\begin{align}
\begin{split}
  \delta_{\hat\scX'}\delta_{\hat\scX} n_\mu
  &= \lie_{\chi'}\lie_\chi n_\mu\,,
  \\
  \delta_{\hat\scX'}\delta_{\hat\scX} e^a_\mu
  &= \lie_{\chi'}\lie_\chi e^a_\mu
    + \lb \lie_\chi\Omega'^a{}_b + \Omega'^a{}_c \Omega^c{}_b \rb e^b_\mu
    \\
  &\qquad
    \textcolor{gray}{-\,\lie_\chi (\Omega'^a{}_b e^b_\mu)
    - \lie_{\chi'}(\Omega^a{}_b e^b_\mu)}\,,
    \\
  \delta_{\hat\scX'}\delta_{\hat\scX} A_\mu
  &= \lie_{\chi'} \lie_\chi A_\mu
    - \dow_\mu \lie_{\chi}\Lambda'
    - e^a_\mu \lb \lie_{\chi}\psi'_a
    + \psi'_b \Omega^b{}_a \rb \,,
    \\
  &\qquad
    \textcolor{gray}{+\,\lie_\chi \dow_\mu \Lambda'
    + \lie_{\chi'} \dow_\mu\Lambda
    + \lie_\chi (e^a_\mu\psi'_a)
    + \lie_{\chi'}(e^a_\mu \psi_a)}\,,
    \\
  \delta_{\hat\scX'}\delta_{\hat\scX} A_{ab}
  &= \lie_{\chi'}\lie_\chi A_{ab}
    - \lb \lie_\chi \Omega'^c{}_a 
    - \Omega^c{}_d \Omega'^d{}_a \rb A_{cb}
    - \lb \lie_\chi \Omega'^c{}_b 
    - \Omega^c{}_d \Omega'^d{}_b \rb A_{ca} 
    \\
  &\qquad
    - e^\mu_a\nabla_\mu \lb \lie_{\chi}\psi'_b + \psi'_c \Omega^c{}_b \rb
    - e^\mu_b\nabla_\mu \lb \lie_{\chi}\psi'_a + \psi'_c \Omega^c{}_a \rb 
    \\
  &\qquad
    \textcolor{gray}{
    +\, e_c^\mu \lb \Omega^c{}_a \nabla_\mu \psi'_b
    + \Omega'^c{}_a \nabla_\mu\psi_b \rb
    + e_b^\mu\lb \Omega^c{}_a \nabla_\mu \psi'_c
    + \Omega'^c{}_a \nabla_\mu\psi_c \rb} 
    \\
  &\qquad
    \textcolor{gray}{
    +\, e_c^\mu \lb \Omega^c{}_b  \nabla_\mu \psi'_a
    + \Omega'^c{}_b \nabla_\mu\psi_a \rb
    + e_a^\mu \lb \Omega^c{}_b \nabla_\mu \psi'_c
    + \Omega'^c{}_b \nabla_\mu\psi_c \rb} 
    \\
  &\qquad
    \textcolor{gray}{
    +\, \lb \Omega'^d{}_b \Omega^c{}_a 
    + \Omega'^d{}_a \Omega^c{}_b \rb A_{cd}
    + \lie_\chi \lb \Omega'^c{}_a A_{cb}
    + \Omega'^c{}_b A_{ac}
    + e^\mu_a\nabla_\mu\psi'_b + e^\mu_b\nabla_\mu\psi'_a \rb} 
    \\
  &\qquad
    \textcolor{gray}{
    +\, \lie_{\chi'} \lb \Omega^c{}_a A_{cb}
    + \Omega^c{}_b A_{ac}
    + e^\mu_a\nabla_\mu\psi_b + e^\mu_b\nabla_\mu\psi_a \rb}\,.
\end{split}
\end{align}
The gray terms are symmetric under the exchange of
$\hat\scX\leftrightarrow\hat\scX'$ and drop out of the commutator. With the
remaining terms, we can see that
$[\delta_{\scX'},\delta_\scX]=\delta_{[\scX',\scX]}$ while acting on the
background fields. Thus the $\delta_{\hat{\scX}}$'s generate a Lie algebra as
claimed.

\subsection{Multipole connections and transformation properties}
\label{app:multipole-connection}

In this Appendix, we derive the explicit form of the objects ${\cal Y}^{(r,s)}$
that appear in the definition of multipole ``connections'' in Section
\ref{sec:multipole-moments}. We begin by postulating that there exists an $r^{\rm th}$
pole connection $A_{(r)}$ that varies under multipole transformations as
follows
\begin{equation}
  A^{\lambda_1\ldots \lambda_{r}}{}_{(r)\mu}
  \to A^{\lambda_1\ldots \lambda_{r}}{}_{(r)\mu}
  + \nabla_\mu \psi_{(r)}^{\lambda_1\ldots \lambda_r}
  + \sum_{s=1}^{r}
  {\cal Y}_\mu^{(r,s)}{}^{\lambda_1\ldots\lambda_{r}}_{\nu_1\ldots \nu_s}
  \psi_{(s)}^{\nu_1\ldots \nu_s} 
  + h_{\mu\nu} \psi^{\nu\lambda_1\ldots\lambda_r}_{(r+1)}\,.
\end{equation}
We can allow $r$ to be zero as well, in which case the connection is just the
monopole gauge field $A_\mu = A_{(0)\mu}$, with the $U(1)$ gauge parameter being
$\psi_{(0)} = \Lambda$. Using these, we can define the $r^{\rm}$ ``field strength''
$F_{(r)}$ as in Eq. \eqref{eq:multipole-connection}, so that it transforms
homogeneously under all the $s$-pole transformations for $s\leq r$. Explicitly
we find
\begin{align}
  F^{\lambda_1\ldots \lambda_{r}}{}_{(r)\mu\nu}
  &\to F^{\lambda_1\ldots \lambda_{r}}{}_{(r)\mu\nu}
    + rR^{(\lambda_1}{}_{\rho\mu\nu} \psi_{(r)}^{\lambda_2\ldots
    \lambda_{r})\rho} 
    + \sum_{s=2}^{r} 
    2{\cal Y}^{(r,s-1)}_{[\mu}{}^{\lambda_1\ldots\lambda_{r}}_{\rho_2\ldots\rho_{s}}
    h_{\nu]\rho_1} 
    \psi_{(s)}^{\rho_1\ldots\rho_{s}} 
    \nn
    \\
  &\qquad
    + \sum_{s=1}^{r} \Bigg[\lb v^{\rho} F^n_{\mu\nu}
    - 2 \delta_{[\mu}^\rho \nabla_{\nu]}
    \rb
    {\cal Y}_\rho^{(r,s)}{}^{\lambda_1\ldots\lambda_{r}}_{\nu_1\ldots \nu_s}
    + \sum_{t=s}^{r} 
    2{\cal Y}^{(r,t)}_{[\mu}{}^{\lambda_1\ldots\lambda_{r}}_{\rho_1\ldots\rho_t}
    {\cal Y}_{\nu]}^{(t,s)}{}^{\rho_1\ldots\rho_{t}}_{\nu_1\ldots \nu_s}
    \Bigg]
    \psi_{(s)}^{\nu_1\ldots \nu_s}  
    \nn
    \\
  &\qquad\qquad
    - 2h_{\rho[\mu}\nabla_{\nu]} \psi_{(r+1)}^{\rho\lambda_1\ldots\lambda_{r}}
    + 2\nabla_{[\mu} h_{\nu]\rho}
    \psi_{(r+1)}^{\rho\lambda_1\ldots \lambda_{r}}
    + 2 {\cal
    Y}^{(r,r)}_{[\mu}{}^{\lambda_1\ldots\lambda_{r}}_{\rho_1\ldots\rho_{r}}
    h_{\nu]\sigma}
    \psi_{(r+1)}^{\sigma\rho_1\ldots\rho_{r}}\,.
\end{align}
The important piece is here is the inhomogeneous term in the third line. Using
$F_{(r)}$ and the spatial $(r+1)^{\rm}$ pole ``connection'' $a_{(r+1)}$ given in
Eq. \eqref{eq:background-variations-quad}, we can construct an object that
transforms like a connection under $\psi_{(r+1)}$ shifts. This is precisely the
definition of $A_{(r+1)}$ given in Eq. \eqref{eq:multipole-connection}.  We
compute its transformation to be
\begin{align}
  A^{\lambda_1\ldots \lambda_{r+1}}{}_{(r+1)\mu}
  &\to A^{\lambda_1\ldots \lambda_{r+1}}{}_{(r+1)\mu}
    + \nabla_{\mu} \psi_{(r+1)}^{\lambda_1\ldots\lambda_{r+1}}
    + \lb n_\mu v^\sigma + \frac{r+1}{r+2} h^\sigma_\mu \rb
    rR^{\lambda_1}{}_{\rho\sigma\nu}
    h^{\lambda_{r+1}\nu}
    \psi_{(r)}^{\lambda_2\ldots\lambda_r\rho} 
    \nn
    \\
  &\qquad
    + \sum_{s=2}^{r+1}
    \lb n_\mu v^\sigma h^{\lambda_{1}}_{\nu_1}
    + \frac{r+1}{r+2} \lb h_\mu^\sigma h^{\lambda_{1}}_{\nu_1}
    - h_{\mu\nu_1} h^{\lambda_{1}\sigma} \rb \rb
    {\cal Y}^{(r,s-1)}_{\sigma}{}^{\lambda_2\ldots\lambda_{r+1}}_{\nu_2\ldots\nu_{s}}
    \psi_{(s)}^{\nu_1\ldots\nu_{s}} 
    \nn
    \\
  &\qquad\qquad 
    + \lb n_\mu v^\sigma + \frac{r+1}{r+2} h^\sigma_\mu \rb
    \sum_{s=1}^{r} \Bigg[\lb v^{\rho} F^n_{\sigma\nu}
    - 2 \delta_{[\sigma}^\rho \nabla_{\nu]} \rb
    {\cal Y}_\rho^{(r,s)}{}^{\lambda_1\ldots\lambda_{r}}_{\nu_1\ldots \nu_s}
    h^{\lambda_{r+1}\nu}
    \\
  &\qquad\qquad\qquad\qquad
    + \sum_{t=s}^{r} 
    2{\cal Y}^{(r,t)}_{[\sigma}{}^{\lambda_1\ldots\lambda_{r}}_{\rho_1\ldots\rho_t}
    {\cal Y}_{\nu]}^{(t,s)}{}^{\rho_1\ldots\rho_{t}}_{\nu_1\ldots \nu_s}
    h^{\lambda_{r+1}\nu}
    \Bigg]
    \psi_{(s)}^{\nu_1\ldots \nu_s}  
    \nn
    \\
  &\qquad \qquad\qquad \qquad  \qquad
    + n_\mu \psi_{(r+1)}^{\rho\lambda_1\ldots \lambda_{r}}
    \nabla_{\rho} v^{\lambda_{r+1}}
  + \psi_{(r+2)}^{\lambda_1\ldots\lambda_{r+1}\rho}h_{\rho\mu}\,.
  \nn
\end{align}
The indices $\lambda_{1}\ldots \lambda_{r+1}$ are understood to be
symmetrised. We can compare this expression to our original expression for
$A_{(r)}$ and write down a series of recursion relations for the
${\cal Y}^{(r,s)}$. The easiest of these are for $s=r$. We find
\begin{align}
\begin{split}
  {\cal Y}^{(r,r)}_{\mu}{}^{\lambda_1\ldots\lambda_{r}}_{\nu_1\ldots\nu_{r}}
  &= n_\mu \nabla_{\nu_1} v^{\lambda_1}
    h^{\lambda_2}_{\nu_2}\ldots h^{\lambda_r}_{\nu_r}
    \\
  &\qquad\qquad
  + \lb n_\mu v^\sigma h^{\lambda_{1}}_{\nu_1}
    + \frac{r}{r+1} \lb h_\mu^\sigma h^{\lambda_{1}}_{\nu_1}
    - h_{\mu\nu_1} h^{\lambda_{1}\sigma} \rb \rb
    {\cal Y}^{(r-1,r-1)}_{\sigma}{}^{\lambda_2\ldots\lambda_{r}}_{\nu_2\ldots\nu_{r}}\,.
\end{split}
\end{align}
The $\lambda_1,\ldots,\lambda_r$ and $\nu_1,\ldots,\nu_r$ are understood to be
symmetrised. This recursion is particularly simple to solve and one simply gets
\begin{equation}
  {\cal Y}^{(r,r)}_{\mu}{}^{\lambda_1\ldots\lambda_{r}}_{\nu_1\ldots\nu_{r}}
  = r\,n_\mu \nabla_{\nu_1} v^{\lambda_1}
    h^{\lambda_2}_{\nu_2}\ldots h^{\lambda_r}_{\nu_r}\,.
\end{equation}
Plugging this result in, we can obtain a recursion relation for $s=r-1$, leading
to
\begin{align}
  {\cal Y}^{(r,r-1)}_{\mu}{}^{\lambda_1\ldots\lambda_{r}}_{\nu_1\ldots\nu_{r-1}}
  &= (r-1)\lb n_\mu v^\sigma + \frac{r}{r+1} h^\sigma_\mu \rb
    {\cal R}^{\lambda_1}{}_{\nu_1\sigma\rho}
    h^{\lambda_2}_{\nu_2}\ldots h^{\lambda_{r-1}}_{\nu_{r-1}}
    h^{\lambda_{r}\rho} \nn\\
  &\qquad
    + \lb n_\mu v^\sigma h^{\lambda_{1}}_{\nu_1}
    + \frac{r}{r+1} \lb h_\mu^\sigma h^{\lambda_{1}}_{\nu_1}
    - h_{\mu\nu_1} h^{\lambda_{1}\sigma} \rb \rb
    {\cal Y}^{(r-1,r-2)}_{\sigma}{}^{\lambda_2\ldots\lambda_{r}}_{\nu_2\ldots\nu_{r-1}},
\end{align}
where ${\cal R}^\lambda{}_{\rho\mu\nu}$ was defined in Eq. \eqref{eq:calR}. We are
unable to solve this recursion explicitly. All further recursion relations are
decisively harder. For $1<s<r-1$, we get
\begin{align}
\begin{split}
  {\cal Y}^{(r,s)}_{\mu}{}^{\lambda_1\ldots\lambda_{r}}_{\nu_1\ldots\nu_{s}}
  &= \lb n_\mu v^\sigma h^{\lambda_{1}}_{\nu_1}
    + \frac{r}{r+1} \lb h_\mu^\sigma h^{\lambda_{1}}_{\nu_1}
    - h_{\mu\nu_1} h^{\lambda_{1}\sigma} \rb \rb
    {\cal Y}^{(r-1,s-1)}_{\sigma}{}^{\lambda_2\ldots\lambda_{r}}_{\nu_2\ldots\nu_{s}} 
    \\
  &\qquad\qquad
    + \lb n_\mu v^\sigma + \frac{r}{r+1} h^\sigma_\mu \rb
    \Bigg[\lb v^{\rho} F^n_{\sigma\nu}
    - 2 \delta_{[\sigma}^\rho \nabla_{\nu]} \rb
    {\cal Y}_\rho^{(r-1,s)}{}^{\lambda_1\ldots\lambda_{r-1}}_{\nu_1\ldots \nu_s}
    h^{\lambda_{r}\nu}
    \\
  &\qquad\qquad\qquad
    + \sum_{t=s}^{r-1} 
    2{\cal Y}^{(r-1,t)}_{[\sigma}{}^{\lambda_1\ldots\lambda_{r-1}}_{\rho_1\ldots\rho_t}
    {\cal Y}_{\nu]}^{(t,s)}{}^{\rho_1\ldots\rho_{t}}_{\nu_1\ldots \nu_s}
    h^{\lambda_{r}\nu}
    \Bigg]\,.
\end{split}
\end{align}
Finally for $s=1$ we have
\begin{align}
\begin{split}
  {\cal Y}^{(r,1)}_{\mu}{}^{\lambda_1\ldots\lambda_r}_{\nu_1}
  &= \lb n_\mu v^\sigma + \frac{r}{r+1} h^\sigma_\mu \rb
    \Bigg[\lb v^{\rho} F^n_{\sigma\nu}
    - 2 \delta_{[\sigma}^\rho \nabla_{\nu]} \rb
    {\cal Y}_\rho^{(r-1,1)}{}^{\lambda_1\ldots\lambda_{r-1}}_{\nu_1\ldots \nu_1}
    h^{\lambda_{r}\nu}
    \\
  &\qquad\qquad\qquad
    + \sum_{t=1}^{r-1} 
    2{\cal Y}^{(r-1,t)}_{[\sigma}{}^{\lambda_1\ldots\lambda_{r-1}}_{\rho_1\ldots\rho_t}
    {\cal Y}_{\nu]}^{(t,1)}{}^{\rho_1\ldots\rho_{t}}_{\nu_1}
    h^{\lambda_{r}\nu}
    \Bigg]\,.
\end{split}
\end{align}

\subsection{Multipole algebra in curved space}
\label{app:algebra}

Let us explicitly compute the commutator of variations
$[\delta_{\hat\scX'},\delta_{\hat\scX}]$ in terms of the generator-decomposition
of $\delta_{\hat\scX}$ given in Eq. \eqref{eq:dB-decomposition}. We find
\begin{align}
  [\delta_{\hat\scX'},\delta_{\hat\scX}] 
  &= \delta_{[\hat\scX',\hat\scX]}
    + \chi^\mu \chi'^\nu \lb
    e^a_\mu e^b_\nu [{\rm P}_a,{\rm P}_b]
    - (n_\mu e^a_\nu - n_\nu e^a_\mu) [{\rm H},{\rm P}_a]
    + i {\cal C}_{\mu\nu} \rb \nn\\
  &\qquad
    + \frac{1}{2}  \lb (\omega^{ab}+\chi^\mu \omega^{ab}{}_\mu) \chi'^\nu n_\nu
    - (\omega'^{ab}+\chi'^\mu \omega^{ab}{}_\mu) \chi^\nu n_\nu \rb
    [{\rm M}_{ab},{\rm H}] \nn\\
  &\qquad
    - \frac{1}{2} \lb 
    (\Omega^{ab}+\chi^\mu\omega^{ab}{}_{\mu}) \chi'^\nu e_\nu^c
    - (\Omega'^{ab}+\chi'^\mu\omega^{ab}{}_{\mu}) \chi^\nu e_\nu^c \rb
    \Big([{\rm M}_{ab},{\rm P}_c] - 2i \delta_{ac} {\rm P}_b \Big) \nn\\
  &\qquad
    + \frac{1}{4} \lb\Omega^{ab} + \chi^\mu \omega^{ab}{}_{\mu} \rb
    \lb\Omega'^{cd} + \chi'^\mu \omega^{cd}{}_{\mu} \rb
    \Big( [{\rm M}_{ab},{\rm M}_{cd}]
    - 4i\delta_{ac} {\rm M}_{bd} \Big)
    \nn\\
  &\qquad
    - \frac{1}{2} \lb
    (\Omega^{ab}+\chi^\mu\omega^{ab}{}_{\mu})(\psi'^c + \chi'^\nu A_\nu^c)
    - (\Omega'^{ab}+\chi'^\mu\omega^{ab}{}_{\mu})(\psi^c + \chi^\nu A_\nu^c)\rb
    \Big( [{\rm M}_{ab},{\rm D}_c] - 2i \delta_{ac} {\rm D}_b \Big) \nn\\
  &\qquad
  \label{eq:generator-commutator}
    - \lb \chi^\mu n_\mu (\Lambda'+\chi'^\nu A_\nu)
    - \chi'^\mu n_\mu (\Lambda+\chi^\nu A_\nu) \rb
    [{\rm H},{\rm Q}] \\
  &\qquad
    + \lb \chi^\mu e_\mu^a \lb \Lambda' + \chi'^\nu A_\nu \rb
    - \chi'^\mu e_\mu^a \lb \Lambda + \chi^\nu A_\nu \rb
    \rb [{\rm P}_a,{\rm Q}] \nn\\
  &\qquad
    - \frac{1}{2}
    \lb (\Omega^{ab} + \chi^\mu \omega^{ab}{}_{\mu})(\Lambda' + \chi'^\nu A_\nu)
    - (\Omega'^{ab} + \chi'^\mu \omega^{ab}{}_{\mu})(\Lambda + \chi^\nu A_\nu)
    \rb [{\rm M}_{ab},{\rm Q}] \nn\\
    &\qquad
    + \lb (\psi^a+\chi^\mu A_\mu^a)(\Lambda' + \chi'^\nu A_\nu)
    - (\psi'^a+\chi'^\mu A_\mu^a)(\Lambda + \chi^\nu A_\nu)
    \rb [{\rm D}_a,{\rm Q}] \nn\\
  &\qquad
    - \lb \chi^\mu n_\mu (\psi'^a + \chi'^\nu A_{\nu}^a)
    - \chi'^\mu n_\mu (\psi^a + \chi^\nu A_{\nu}^a)
    \rb \Big([{\rm H},{\rm D}_a]
    + \frac{i}{2} e_a^{\rho}e^{b\sigma}\lie_v h_{\rho\sigma} {\rm D}_b\Big) \nn\\
  &\qquad
    + \lb \chi^\mu e_\mu^a (\psi'^b + \chi'^\nu A^b_{\nu})
    - \chi'^\mu e_\mu^a (\psi^b + \chi^\nu A^b_{\nu})\rb
    \Big([{\rm P}_a,{\rm D}_b] -i\delta_{ab}{\rm Q} \Big) \nn\\
  &\qquad
    + (\psi^a+\chi^\mu A^a_\mu)(\psi'^b+\chi'^\nu A^b_\nu) [{\rm D}_a,{\rm
    D}_b]\,.
    \nn
\end{align}
The curvature  ${\cal C}_{\mu\nu}$ is defined in
Eq. \eqref{eq:curvature-operator}. Requiring the RHS of
Eq. \eqref{eq:generator-commutator} to be just $\delta_{[\hat\scX',\hat\scX]}$,
we can read out the commutation relations of the curved space dipole algebra
given in Eq. \eqref{eq:curved-dipole-algebra}. This computation can be
generalised to higher multipole moments as well.

\section{Background coupling of scalar charge theory}

The covariant Lagrangian is given as
\begin{align}
  \mathcal{L}
  &= 
    \frac{i}{2}
    \lb \Phi^* v^\mu \Df_\mu\Phi
    - \Phi v^\mu \Df_\mu\Phi^* \rb
    - \lambda_S
    \Big( h^{\mu\nu}\Df_{\mu\nu}(\Phi^*,\Phi^*) + \gamma {\Phi^*}^2 \Big)
    \Big(h^{\rho\sigma}\Df_{\rho\sigma}(\Phi,\Phi) + \gamma \Phi^2 \Big) \nn\\
  &\qquad\qquad\qquad\qquad
  - \lambda_T h^{\mu\rho} h^{\nu\sigma} \Df_{\mu\nu}(\Phi^*,\Phi^*)
  \Df_{\rho\sigma}(\Phi,\Phi)
  - V(\Phi^*\Phi).
\end{align}
The non-local covariant derivatives were defined before. The variation of the
Lagrangian with respect to the background fields is given as
\begin{align}
  \frac{1}{\sqrt{\gamma}}\delta(\sqrt{\gamma}\,\mathcal{L})
  &=
    \Big( \mathcal{L} v^\mu
    - \frac{i}{2} \lb \Phi^* v^\mu v^\nu\Df_\nu\Phi
    - \Phi v^\mu v^\nu\Df_\nu\Phi^*
    \rb
    + 2\mathcal{B}_{\rho\nu} v^\rho h^{\mu\nu}
    \Big) \delta n_\mu  \nn\\
  &\qquad
    + \lb \half \mathcal{L} h^{\mu\nu} 
    - \frac{i}{2}
    \lb \Phi^* v^\mu h^{\nu\rho}\Df_\rho\Phi
    - \Phi v^\mu h^{\nu\rho}\Df_\rho\Phi^*
    \rb
    + \mathcal{B}_{\rho\sigma} h^{\rho\mu} h^{\sigma\nu} \rb \delta h_{\mu\nu}
    + q\Phi^*\Phi\, v^\mu\delta A_\mu
    \nn\\
  &\qquad
    - \mathcal{A}^{*\mu\nu} \delta\mathcal{X}_{\mu\nu}
    - \mathcal{A}^{\mu\nu} \delta\mathcal{X}^*_{\mu\nu},
\end{align}
where we have defined 
\begin{align}
  \mathcal{X}_{\mu\nu}
  &= \Phi\,\Df_{(\mu} \Df_{\nu)}\Phi
  - \Df_\mu\Phi\,\Df_\nu\Phi
  + \half iq\Phi^2 a_{\mu\nu}, \nn\\
  \mathcal{A}^{\mu\nu}
  &= \lambda_S
    \Big( h^{\rho\sigma}\mathcal{X}_{\rho\sigma} + \gamma {\Phi}^2 \Big)
    h^{\mu\nu}
    + \lambda_T h^{\mu\rho} h^{\nu\sigma}\mathcal{X}_{\rho\sigma}, \nn\\
  \mathcal{B}_{\mu\nu}
  &= \lambda_S
    \Big( h^{\rho\sigma}\mathcal{X}^*_{\rho\sigma} + \gamma {\Phi^*}^2 \Big)
    \mathcal{X}_{\mu\nu}
    + \lambda_S
    \Big( h^{\rho\sigma}\mathcal{X}_{\rho\sigma} + \gamma {\Phi}^2 \Big)
    \mathcal{X}^*_{\mu\nu}
    + 2\lambda_T h^{\rho\sigma} \mathcal{X}^*_{\rho(\mu}\mathcal{X}_{\nu)\sigma}.
\end{align}
All the technicalities from the variation are captured in the variation of
$\mathcal{X}_{\mu\nu}$; we get
\begin{align}
  \mathcal{A}^{*\mu\nu}\delta\mathcal{X}_{\mu\nu}
  &= \half iq\Phi^2 \mathcal{A}^{*\mu\nu} \delta a_{\mu\nu}
    -iq\Phi^2\,\mathcal{A}^{*\mu\nu} \nabla_{\mu}\delta A_{\nu} \nn\\
  &\qquad
    - \Phi \Df_\lambda\Phi \bigg(
    v^\lambda \mathcal{A}^{*\mu\nu} \nabla_{\mu} \delta n_{\nu}
    + \half \left(
    2h^{\lambda(\mu} \mathcal{A}^{*\nu)\rho} 
    - h^{\lambda\rho} \mathcal{A}^{*\mu\nu} 
  \right) \nabla_\rho \delta h_{\mu\nu} \\
  &\qquad\qquad\qquad
    + \half h^{\lambda\rho} \mathcal{A}^{*\mu\nu} (\lie_v h_{\mu\nu} )\delta n_\rho
    - v^\mu h^{\lambda\sigma} F^n_{\sigma\rho} \mathcal{A}^{*\rho\nu}
     \delta h_{\mu\nu} \bigg),
\end{align}
where we have used the variation of the connection given in
Eq.~\eqref{eq:connection-variation}. After ignoring some total derivative terms,
we get
\begin{align}
  \mathcal{A}^{*\mu\nu}\delta\mathcal{X}_{\mu\nu}
  &= - \lb
    \half h^{\lambda\mu} \lie_v h_{\rho\sigma}
    \mathcal{A}^{*\rho\sigma}  \Phi \Df_\lambda\Phi 
    - \nabla'_\nu \lb \Phi\, v^\lambda \Df_\lambda\Phi
    \mathcal{A}^{*\mu\nu}\rb
    \rb
    \delta n_{\mu} \nn\\ 
  &\qquad
    + \lb v^\mu h^{\lambda\sigma} F^n_{\sigma\rho} \mathcal{A}^{*\rho\nu}
    \Phi \Df_\lambda\Phi
    + \half \nabla'_\rho \lb \left(
    2h^{\lambda(\mu} \mathcal{A}^{*\nu)\rho} 
    - h^{\lambda\rho} \mathcal{A}^{*\mu\nu} 
    \right) \Phi \Df_\lambda\Phi \rb \rb \delta h_{\mu\nu} \nn\\
  &\qquad
    + \nabla'_\nu \lb iq\mathcal{A}^{*\mu\nu}\Phi^2 \rb \delta A_{\mu}
    + \half iq\mathcal{A}^{*\mu\nu}\Phi^2 \delta a_{\mu\nu}.
\end{align}
Using these, we can read out the conserved currents. The $U(1)$ monopole and
dipole currents are given simply as
\begin{align}
  J^\mu
  &= q\Phi^*\Phi\, v^\mu
    - \nabla'_\nu \lb iq\mathcal{A}^{*\mu\nu}\Phi^2
    - iq{\Phi^*}^2 \mathcal{A}^{\mu\nu}\rb, \nn\\
  J^{\mu\nu}
  &= - iq\mathcal{A}^{*\mu\nu}\Phi^2
    + iq{\Phi^*}^2\mathcal{A}^{\mu\nu}.
\end{align}
On the other hand, the energy current, momentum density, and stress tensor are
given as
\begin{align}
  \epsilon^\mu 
  &= v^\mu \lb \frac{i}{2}\lb \Phi^* v^\nu\Df_\nu\Phi - \Phi v^\nu\Df_\nu\Phi^* \rb
    - \mathcal{L}  \rb
    - 2\mathcal{B}_{\rho\nu} v^\rho h^{\mu\nu}  \nn\\
  &\qquad
    + \nabla'_\nu \lb
    \mathcal{A}^{*\mu\nu}\Phi\, v^\lambda \Df_\lambda\Phi 
    + \Phi^* v^\lambda \Df_\lambda\Phi^* \mathcal{A}^{\mu\nu}
    \rb  \nn\\
  &\qquad
    - h^{\lambda\mu}
    \lb \mathcal{A}^{*\rho\sigma}  \Phi \Df_\lambda\Phi
    + \Phi^* \Df_\lambda\Phi^* \mathcal{A}^{\rho\sigma} 
    \rb
    \half\lie_v h_{\rho\sigma}
    - iq\lb \mathcal{A}^{*\mu\nu}\Phi^2
    - {\Phi^*}^2\mathcal{A}^{\mu\nu} \rb v^\rho F_{\rho\nu}, \nn\\
  \pi^\mu
  &=  - \frac{i}{2} \lb
    \Phi^* h^{\mu\nu}\Df_\nu\Phi
    - \Phi h^{\mu\nu}\Df_\nu\Phi^*
    \rb
    - h^{\lambda\sigma} F^n_{\sigma\rho}
    \lb \mathcal{A}^{*\rho\mu} \Phi \Df_\lambda\Phi
    +  \Phi^* \Df_\lambda\Phi^* \mathcal{A}^{\rho\mu} \rb, \nn\\
  \tau^{\mu\nu} + \tau^{\mu\nu}_{\text d}
  &= \mathcal{L} h^{\mu\nu} 
    + 2\mathcal{B}_{\rho\sigma} h^{\rho\mu} h^{\sigma\nu} \nn\\
  &\qquad
    - \nabla'_\rho \lb \left(
    2h^{\lambda(\mu} \mathcal{A}^{*\nu)\rho} 
    - h^{\lambda\rho} \mathcal{A}^{*\mu\nu} 
    \right) \Phi \Df_\lambda\Phi
    + \Phi^* \Df_\lambda\Phi^* \left(
    2h^{\lambda(\mu} \mathcal{A}^{\nu)\rho} 
    - h^{\lambda\rho} \mathcal{A}^{\mu\nu} 
    \right) \rb \nn\\
  &\qquad
    - iq\lb \mathcal{A}^{*\mu\rho}\Phi^2
    - {\Phi^*}^2\mathcal{A}^{\mu\rho}
    \rb A^{\nu}{}_\rho,
\end{align}

The expressions are considerably simpler in flat space when all the background
sources have been switched off. For the $U(1)$ monopole and dipole currents, we
revert back to the expressions we found before
\begin{align}
  J^t
  &= q\Phi^*\Phi, \nn\\
  J^i
  &= - \dow_j \lb iq\mathcal{A}^{*ij}\Phi^2
    - iq{\Phi^*}^2 \mathcal{A}^{ij}\rb, \nn\\
  J^{ij}
  &= - iq\mathcal{A}^{*ij}\Phi^2
    + iq{\Phi^*}^2\mathcal{A}^{ij},
\end{align}
whereas for the energy density, energy current, momentum density, and stress
tensor, we get
\begin{align}
  \epsilon^t
  &= \frac{i}{2}\lb \Phi^*\dow_t\Phi - \Phi\dow_t\Phi^* \rb
    - \mathcal{L}  , \nn\\
  \epsilon^i
  &= - 2\mathcal{B}^i_{t}
    + \dow_j \lb
    \mathcal{A}^{*ij}\Phi\dow_t\Phi 
    + \Phi^* \dow_t \Phi^* \mathcal{A}^{ij}
    \rb , \nn\\
  \pi^i
  &=  - \frac{i}{2} \lb \Phi^*\dow^i\Phi - \Phi\dow^i\Phi^* \rb, \nn\\
  \tau^{ij} 
  &= \mathcal{L} \delta^{ij}
    + 2\mathcal{B}^{ij} 
    - \dow_k \lb 
    2\mathcal{A}^{*k(i} \Phi \dow^{j)}\Phi
    - \mathcal{A}^{*ij} \Phi \dow^k \Phi
    + 2\Phi^* \dow^{(i}\Phi^* \mathcal{A}^{j)k} 
    - \Phi^* \dow^k \Phi^* \mathcal{A}^{ij}\rb ,
\end{align}
where $\mathcal{A}^{ij}$, $\mathcal{B}_{ij}$, and $\mathcal{B}_{ti}$ are defined as
\begin{align}
  \mathcal{X}_{ti}
  &= \Phi\,\dow_i \dow_t \Phi
  - \dow_i\Phi\,\dow_t\Phi, \nn\\
  \mathcal{X}_{ij}
  &= \Phi\,\dow_i \dow_j \Phi
  - \dow_i\Phi\,\dow_j\Phi, \nn\\
  \mathcal{A}^{ij}
  &= \lambda_S
    \Big(\mathcal{X}^k{}_{k} + \gamma {\Phi}^2 \Big)
    \delta^{ij}
    + \lambda_T \mathcal{X}^{ij}, \nn\\
  \mathcal{B}_{ti}
  &= \lambda_S
    (\mathcal{X}^k{}_k + \gamma {\Phi}^2)^*\mathcal{X}_{ti}
    + \lambda_S
    (\mathcal{X}^k{}_k + \gamma {\Phi}^2)\mathcal{X}^*_{ti}
    + \lambda_T h^{\rho\sigma}
    \lb \mathcal{X}^*_{tk}\mathcal{X}_{i}{}^k
    + \mathcal{X}^*_{ki}\mathcal{X}_{t}{}^k \rb, \nn\\
  \mathcal{B}_{ij}
  &= \lambda_S
    (\mathcal{X}^k{}_k + \gamma {\Phi}^2)^*\mathcal{X}_{ij}
    + \lambda_S
    (\mathcal{X}^k{}_k + \gamma {\Phi}^2)\mathcal{X}^*_{ij}
    + 2\lambda_T \mathcal{X}^*_{k(i}\mathcal{X}_{j)}{}^k.
\end{align}
Note that $\tau^{ij}_{\mathrm d}$ is identically zero in the absence of
background fields.

\bibliography{refs}

\providecommand{\href}[2]{#2}\begingroup\raggedright\begin{thebibliography}{10}

\bibitem{Chamon:2004lew}
C.~Chamon, {\it {Quantum Glassiness}},  {\em Phys. Rev. Lett.} {\bf 94} (2005),
  no.~4 040402, \href{http://xxx.lanl.gov/abs/cond-mat/0404182}{{\tt
  cond-mat/0404182}}.

\bibitem{2011AnPhy.326..839B}
S.~{Bravyi}, B.~{Leemhuis}, and B.~M. {Terhal}, {\it {Topological order in an
  exactly solvable 3D spin model}},  {\em Annals of Physics} {\bf 326} (Apr.,
  2011) 839--866, \href{http://xxx.lanl.gov/abs/1006.4871}{{\tt 1006.4871}}.

\bibitem{Haah:2011drr}
J.~Haah, {\it {Local stabilizer codes in three dimensions without string
  logical operators}},  {\em Phys. Rev. A} {\bf 83} (2011), no.~4 042330,
  \href{http://xxx.lanl.gov/abs/1101.1962}{{\tt 1101.1962}}.

\bibitem{Vijay:2015mka}
S.~Vijay, J.~Haah, and L.~Fu, {\it {A New Kind of Topological Quantum Order: A
  Dimensional Hierarchy of Quasiparticles Built from Stationary Excitations}},
  {\em Phys. Rev. B} {\bf 92} (2015), no.~23 235136,
  \href{http://xxx.lanl.gov/abs/1505.02576}{{\tt 1505.02576}}.

\bibitem{Nandkishore:2018sel}
R.~M. Nandkishore and M.~Hermele, {\it {Fractons}},  {\em Ann. Rev. Condensed
  Matter Phys.} {\bf 10} (2019) 295--313,
  \href{http://xxx.lanl.gov/abs/1803.11196}{{\tt 1803.11196}}.

\bibitem{Vijay:2016phm}
S.~Vijay, J.~Haah, and L.~Fu, {\it {Fracton Topological Order, Generalized
  Lattice Gauge Theory and Duality}},  {\em Phys. Rev. B} {\bf 94} (2016),
  no.~23 235157, \href{http://xxx.lanl.gov/abs/1603.04442}{{\tt 1603.04442}}.

\bibitem{Seiberg:2020bhn}
N.~Seiberg and S.-H. Shao, {\it {Exotic Symmetries, Duality, and Fractons in
  2+1-Dimensional Quantum Field Theory}},  {\em SciPost Phys.} {\bf 10} (2021),
  no.~2 027, \href{http://xxx.lanl.gov/abs/2003.10466}{{\tt 2003.10466}}.

\bibitem{Seiberg:2020wsg}
N.~Seiberg and S.-H. Shao, {\it {Exotic $U(1)$ Symmetries, Duality, and
  Fractons in 3+1-Dimensional Quantum Field Theory}},  {\em SciPost Phys.} {\bf
  9} (2020), no.~4 046, \href{http://xxx.lanl.gov/abs/2004.00015}{{\tt
  2004.00015}}.

\bibitem{Williamson:2016jiq}
D.~J. Williamson, {\it {Fractal symmetries: Ungauging the cubic code}},  {\em
  Phys. Rev. B} {\bf 94} (2016), no.~15 155128,
  \href{http://xxx.lanl.gov/abs/1603.05182}{{\tt 1603.05182}}.

\bibitem{You:2018oai}
Y.~You, T.~Devakul, F.~J. Burnell, and S.~L. Sondhi, {\it {Subsystem symmetry
  protected topological order}},  {\em Phys. Rev. B} {\bf 98} (2018), no.~3
  035112, \href{http://xxx.lanl.gov/abs/1803.02369}{{\tt 1803.02369}}.

\bibitem{Gromov:2020yoc}
A.~Gromov, A.~Lucas, and R.~M. Nandkishore, {\it {Fracton hydrodynamics}},
  {\em Phys. Rev. Res.} {\bf 2} (2020), no.~3 033124,
  \href{http://xxx.lanl.gov/abs/2003.09429}{{\tt 2003.09429}}.

\bibitem{Grosvenor:2021rrt}
K.~T. Grosvenor, C.~Hoyos, F.~Pe\~na Benitez, and P.~Sur\'owka, {\it
  {Hydrodynamics of ideal fracton fluids}},
  \href{http://xxx.lanl.gov/abs/2105.01084}{{\tt 2105.01084}}.

\bibitem{Glorioso:2021bif}
P.~Glorioso, J.~Guo, J.~F. Rodriguez-Nieva, and A.~Lucas, {\it {Breakdown of
  hydrodynamics below four dimensions in a fracton fluid}},
  \href{http://xxx.lanl.gov/abs/2105.13365}{{\tt 2105.13365}}.

\bibitem{Pena-Benitez:2021ipo}
F.~Pe\~na Benitez, {\it {Fractons, Symmetric Gauge Fields and Geometry}},
  \href{http://xxx.lanl.gov/abs/2107.13884}{{\tt 2107.13884}}.

\bibitem{Nguyen:2020yve}
D.~X. Nguyen, A.~Gromov, and S.~Moroz, {\it {Fracton-elasticity duality of
  two-dimensional superfluid vortex crystals: defect interactions and quantum
  melting}},  {\em SciPost Phys.} {\bf 9} (2020) 076,
  \href{http://xxx.lanl.gov/abs/2005.12317}{{\tt 2005.12317}}.

\bibitem{Doshi:2020jso}
D.~Doshi and A.~Gromov, {\it {Vortices and Fractons}},
  \href{http://xxx.lanl.gov/abs/2005.03015}{{\tt 2005.03015}}.

\bibitem{2018PhRvL.120s5301P}
M.~{Pretko} and L.~{Radzihovsky}, {\it {Fracton-Elasticity Duality}},  {\em
  Phys.Rev.Lett.} {\bf 120} (May, 2018) 195301,
  \href{http://xxx.lanl.gov/abs/1711.11044}{{\tt 1711.11044}}.

\bibitem{Du:2021pbc}
Y.-H. Du, U.~Mehta, D.~X. Nguyen, and D.~T. Son, {\it {Volume-preserving
  diffeomorphism as nonabelian higher-rank gauge symmetry}},
  \href{http://xxx.lanl.gov/abs/2103.09826}{{\tt 2103.09826}}.

\bibitem{sous2019}
J.~Sous and M.~Pretko, {\it Fractons from polarons},  {\em Physical Review B}
  {\bf 102} (Dec, 2020).

\bibitem{Sous:2020ypq}
J.~Sous and M.~Pretko, {\it {Fractons from frustration in hole-doped
  antiferromagnets}},  {\em Materials} {\bf 5} (2020) 81,
  \href{http://xxx.lanl.gov/abs/2009.05577}{{\tt 2009.05577}}.

\bibitem{Pretko:2016kxt}
M.~Pretko, {\it {Subdimensional Particle Structure of Higher Rank U(1) Spin
  Liquids}},  {\em Phys. Rev. B} {\bf 95} (2017), no.~11 115139,
  \href{http://xxx.lanl.gov/abs/1604.05329}{{\tt 1604.05329}}.

\bibitem{Gromov:2017vir}
A.~Gromov, {\it {Chiral Topological Elasticity and Fracton Order}},  {\em Phys.
  Rev. Lett.} {\bf 122} (2019), no.~7 076403,
  \href{http://xxx.lanl.gov/abs/1712.06600}{{\tt 1712.06600}}.

\bibitem{Slagle:2018kqf}
K.~Slagle, A.~Prem, and M.~Pretko, {\it {Symmetric Tensor Gauge Theories on
  Curved Spaces}},  {\em Annals Phys.} {\bf 410} (2019) 167910,
  \href{http://xxx.lanl.gov/abs/1807.00827}{{\tt 1807.00827}}.

\bibitem{LargeN}
K.~Jensen and A.~Raz, {\it {Large $N$ fractons}},
  \href{http://xxx.lanl.gov/abs/2112.XXXXX}{{\tt 2112.XXXXX}}.

\bibitem{Banerjee:2012iz}
N.~Banerjee, J.~Bhattacharya, S.~Bhattacharyya, S.~Jain, S.~Minwalla, and
  T.~Sharma, {\it {Constraints on Fluid Dynamics from Equilibrium Partition
  Functions}},  {\em JHEP} {\bf 09} (2012) 046,
  \href{http://xxx.lanl.gov/abs/1203.3544}{{\tt 1203.3544}}.

\bibitem{Jensen:2012jh}
K.~Jensen, M.~Kaminski, P.~Kovtun, R.~Meyer, A.~Ritz, and A.~Yarom, {\it
  {Towards hydrodynamics without an entropy current}},  {\em Phys. Rev. Lett.}
  {\bf 109} (2012) 101601, \href{http://xxx.lanl.gov/abs/1203.3556}{{\tt
  1203.3556}}.

\bibitem{Crossley:2015evo}
M.~Crossley, P.~Glorioso, and H.~Liu, {\it {Effective field theory of
  dissipative fluids}},  {\em JHEP} {\bf 09} (2017) 095,
  \href{http://xxx.lanl.gov/abs/1511.03646}{{\tt 1511.03646}}.

\bibitem{Haehl:2015uoc}
F.~M. Haehl, R.~Loganayagam, and M.~Rangamani, {\it {Topological sigma models
  \& dissipative hydrodynamics}},  {\em JHEP} {\bf 04} (2016) 039,
  \href{http://xxx.lanl.gov/abs/1511.07809}{{\tt 1511.07809}}.

\bibitem{Jensen:2017kzi}
K.~Jensen, N.~Pinzani-Fokeeva, and A.~Yarom, {\it {Dissipative hydrodynamics in
  superspace}},  {\em JHEP} {\bf 09} (2018) 127,
  \href{http://xxx.lanl.gov/abs/1701.07436}{{\tt 1701.07436}}.

\bibitem{Jensen:2014ama}
K.~Jensen, {\it {Aspects of hot Galilean field theory}},  {\em JHEP} {\bf 04}
  (2015) 123, \href{http://xxx.lanl.gov/abs/1411.7024}{{\tt 1411.7024}}.

\bibitem{Geracie:2015xfa}
M.~Geracie, K.~Prabhu, and M.~M. Roberts, {\it {Fields and fluids on curved
  non-relativistic spacetimes}},  {\em JHEP} {\bf 08} (2015) 042,
  \href{http://xxx.lanl.gov/abs/1503.02680}{{\tt 1503.02680}}.

\bibitem{Novak:2019wqg}
I.~Novak, J.~Sonner, and B.~Withers, {\it {Hydrodynamics without boosts}},
  {\em JHEP} {\bf 07} (2020) 165,
  \href{http://xxx.lanl.gov/abs/1911.02578}{{\tt 1911.02578}}.

\bibitem{deBoer:2020xlc}
J.~de~Boer, J.~Hartong, E.~Have, N.~A. Obers, and W.~Sybesma, {\it {Non-Boost
  Invariant Fluid Dynamics}},  {\em SciPost Phys.} {\bf 9} (2020), no.~2 018,
  \href{http://xxx.lanl.gov/abs/2004.10759}{{\tt 2004.10759}}.

\bibitem{Armas:2020mpr}
J.~Armas and A.~Jain, {\it {Effective field theory for hydrodynamics without
  boosts}},  {\em SciPost Phys.} {\bf 11} (2021), no.~3 054,
  \href{http://xxx.lanl.gov/abs/2010.15782}{{\tt 2010.15782}}.

\bibitem{Bidussi:2021}
L.~Bidussi, J.~Hartong, E.~Have, J.~Musaeus, and S.~Prohazka, {\it {Fractons,
  dipole symmetries and curved spacetime}},
  \href{http://xxx.lanl.gov/abs/2111.XXXXX}{{\tt 2111.XXXXX}}.

\bibitem{Pretko:2018jbi}
M.~Pretko, {\it {The Fracton Gauge Principle}},  {\em Phys. Rev. B} {\bf 98}
  (2018), no.~11 115134, \href{http://xxx.lanl.gov/abs/1807.11479}{{\tt
  1807.11479}}.

\bibitem{Geracie:2014nka}
M.~Geracie, D.~T. Son, C.~Wu, and S.-F. Wu, {\it {Spacetime Symmetries of the
  Quantum Hall Effect}},  {\em Phys. Rev. D} {\bf 91} (2015) 045030,
  \href{http://xxx.lanl.gov/abs/1407.1252}{{\tt 1407.1252}}.

\bibitem{Jensen:2014aia}
K.~Jensen, {\it {On the coupling of Galilean-invariant field theories to curved
  spacetime}},  {\em SciPost Phys.} {\bf 5} (2018), no.~1 011,
  \href{http://xxx.lanl.gov/abs/1408.6855}{{\tt 1408.6855}}.

\bibitem{Hartong:2014pma}
J.~Hartong, E.~Kiritsis, and N.~A. Obers, {\it {Schr\"odinger Invariance from
  Lifshitz Isometries in Holography and Field Theory}},  {\em Phys. Rev. D}
  {\bf 92} (2015) 066003, \href{http://xxx.lanl.gov/abs/1409.1522}{{\tt
  1409.1522}}.

\bibitem{Jain:2018jxj}
A.~Jain, {\em {A universal framework for hydrodynamics}}.
\newblock PhD thesis, Durham U., CPT, 6, 2018.

\bibitem{Gromov:2018nbv}
A.~Gromov, {\it {Towards classification of Fracton phases: the multipole
  algebra}},  {\em Phys. Rev. X} {\bf 9} (2019), no.~3 031035,
  \href{http://xxx.lanl.gov/abs/1812.05104}{{\tt 1812.05104}}.

\bibitem{Gorantla:2021bda}
P.~Gorantla, H.~T. Lam, N.~Seiberg, and S.-H. Shao, {\it {The low-energy limit
  of some exotic lattice theories and UV/IR mixing}},
  \href{http://xxx.lanl.gov/abs/2108.00020}{{\tt 2108.00020}}.

\bibitem{Burnell:2021reh}
F.~J. Burnell, T.~Devakul, P.~Gorantla, H.~T. Lam, and S.-H. Shao, {\it
  {Anomaly Inflow for Subsystem Symmetries}},
  \href{http://xxx.lanl.gov/abs/2110.09529}{{\tt 2110.09529}}.

\bibitem{Yamaguchi:2021xeq}
S.~Yamaguchi, {\it {Gapless edge modes in (4+1)-dimensional topologically
  massive tensor gauge theory and anomaly inflow for subsystem symmetry}},
  \href{http://xxx.lanl.gov/abs/2110.12861}{{\tt 2110.12861}}.

\end{thebibliography}\endgroup
\bibliographystyle{JHEP}

\end{document}